\documentclass[aps,prx,amsmath,amssymb,reprint,twocolumn, linenumbers]{revtex4-2}

\usepackage[utf8]{inputenc}
\usepackage{graphicx}
\usepackage{tikz}
\usepackage{pgfplots}
\pgfplotsset{compat=newest,legend style={font=\footnotesize},
             ticklabel style={font=\footnotesize},
             x label style={font=\footnotesize},
             y label style={font=\footnotesize}}
\usepgfplotslibrary{groupplots,dateplot}
\usetikzlibrary{patterns,shapes.arrows}

\usepackage{mhchem}

\usepackage{comment}
\usepackage{amsmath,amssymb,amsthm,mathtools,amsfonts}
\usepackage{bm}
\usepackage{algorithm,algorithmic}
\usepackage{bbm}
\usepackage{url}
\usepackage{array}
\usepackage{hyperref}
\usepackage{enumerate}
\usepackage[squaren]{SIunits}
\usepackage{cases}
\usepackage{bbold}
\usepackage{accents}
\usepackage{scalerel}
\usepackage{appendix}
\usepackage{xspace}

\usepackage[commandnameprefix=ifneeded]{changes}
\makeatletter
\@namedef{Changes@AuthorColor}{magenta}
\colorlet{Changes@Color}{magenta}
\makeatother

\makeatletter
\renewcommand{\todo}[2][]{\tikzexternaldisable\@todo[#1]{#2}\tikzexternalenable}
\makeatother

\let\oldv\v

\def\hardouin{hardouin_active_2022}
\def\u{\mathbf{u}}
\def\v{\mathbf{v}}
\def\g{{\mathbf{v}_{\scaleto{\partial\Omega}{4pt}}}}
\def\f{\mathbf{f}}
\def\n{\mathbf{n}}
\def\stress{\boldsymbol{\tau}}
\def\stressadj{\boldsymbol{\sigma}}

\def\rey{\text{Re}}
\def\act{\xi}
\def\visc{\mu}
\def\nemt{\mathrm{\mathbf{Q}}}
\def\nemv{\textbf{q}}
\def\source{r}
\def\fluo{\psi}
\def\magn{a}
\def\cauchy{\boldsymbol{\varsigma}}
\def\inboundary{\partial \mathrm{K}}
\def\indomain{\mathrm{K}}
\def\vcm{\boldsymbol{\nu}}
\def\trac{\boldsymbol{\zeta}}
\def\sm{\mu_s}
\newcommand{\appx}{Appendix\xspace}
\newcommand{\suppx}{Appendix\xspace}
\newcommand{\suppxfig}{Figure\xspace}

\def\mum{\micro\metre}

\newcommand\change[1]{\textcolor{black}{#1}}

\def\v{\mathbf{v}}

\def\x{\mathbf{x}}

\def\n{\mathbf{n}}

\def\y{\mathbf{y}}

\DeclarePairedDelimiterX{\inp}[2]{\langle}{\rangle}{#1, #2}
\newcommand{\norm}[1]{\left\lVert#1\right\rVert}

\DeclarePairedDelimiterX{\inner}[2]{\langle}{\rangle}{#1, #2}
\def\reals{\mathbb{R}}

\def\define{:=}

\renewcommand{\selectlanguage}[1]{}
\begin{document}
\nolinenumbers
\title{Inverse Measurements in Active Nematics}

\author{Aleix Boquet-Pujadas$^{1}$}
\altaffiliation{Corresponding author: aleix.boquetipujadas@epfl.ch}
\author{J\'er\^ome Hardou\"in$^{2}$}
\author{Junhao Wen$^{3}$}
\author{Jordi Ign\'es-Mullol$^{2}$}
\author{Francesc Sagu\'es$^{2}$}

\affiliation{%
 $^1$ Bioimage Analysis Unit, Institut Pasteur, Paris, France\\ Biomedical Imaging Group, École polytechnique fédérale de Lausanne, Lausanne, Switzerland\\
 $^2$ Departament de Qu\'imica F\'isica, Universitat de Barcelona, Barcelona, Spain \\
 Institute of Nanoscience and Nanotechnology, Universitat de Barcelona, Barcelona, Spain \\
 $^3$ Laboratory of AI and Biomedical Science (LABS), University of Southern California, Los Angeles, California, USA
}

\begin{abstract}
    We present a framework to take new measurements in nematic systems that contain active elements such as molecular motors. Spatio-temporal fields of stress, traction, velocity, pressure, and forces are estimated jointly from microscopy images alone. Our inverse-problem approach ensures that these fields comply with physical laws and are accurate at system boundaries. 
Our novel measurements in active biological materials provide new insight for the design of boundary-aware nematic systems. The shear stress at the wall unveils a correlation with \mbox{the nucleation} of topological defects. The pressure gradients and the velocity characterize how boundary effects drive system-wide dynamics. %
And the forces link the underlying fluid with the \mbox{nematic tensor} to reveal the time scales of the system.
More broadly, our work establishes a generalizable approach to quantitatively study experimental systems that are inaccessible to measuring probes.

\end{abstract}
\maketitle

\section{Introduction}

\begin{figure}[!ht]
\centering
    \vspace{-8pt}
    \hspace{-4pt}
    \includegraphics[width=0.48\textwidth]{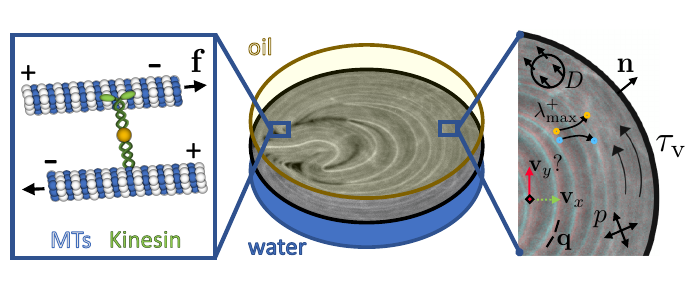}
    \vspace{-14pt}
	\caption{\textbf{Microtubule-kinesin systems as an example of active nematics.} Along with microtubules (MTs) and kinesin, the system contains ATP as an energy source, as well as depleting agents, and streptavidin to cluster the motors into pairs \cite{sanchez_spontaneous_2012}. The system is made to lie at a water-oil interface as a quasi-2D layer. MTs are labelled with a fluorescent compound for imaging. 
 The pairs of kinesin motors pull on the MTs in a direction that is dependent on MT polarity: two adjacent MTs are made to slide past each other if they have opposite polarities, but there is no relative motion otherwise. 
The local sorting results in a global extensional motion that builds higher-scale order. 
This competes with the disintegration rate of the MT bundles to create the dynamical~system. The system is unstable because the active stress amplifies bending deformations until fracture. The resulting advection appears chaotic---with jets and swirls---and leads to the creation and annihilation of $\pm 1/2$ topological defects in the macroscopic orientation field of the MTs \cite{giomi_defect_2013}. The rightmost image illustrates the variables used throughout the manuscript.  
At the red diamond, estimating $\v_x$ is easier than $\v_y$ because the former is perpendicular to the direction $\nemv$ of homogeneous intensity, which aligns with the MTs. 
Conservation equation \eqref{eq:conservation} is represented by $D$, and $\lambda^+_\text{max}$ illustrates Lyapunov exponents using two particles. The background is from Figure~\ref{fig:piv}.a.i.
}
    \label{fig:schematic}
    \vspace{-5.5pt}
\end{figure}

The properties of nematic fluids change drastically when their constitutive elements extract energy from their surroundings and turn it into mechanical work. 
The energy input maintains the system out of equilibrium and shortens the range of the orientational order. This gives rise to emergent, collective motion, sometimes in the form of chaotic flows with motile topological defects~\cite{giomi_geometry_2015,martinez-prat_scaling_2021}.
Many biological materials such as bacterial colonies or epithelial tissues can be regarded as active nematic (AN) systems \cite{duclos_spontaneous_2018,volfson_biomechanical_2008,saw_topological_2017, ilina_cellcell_2020, genkin2017topological}. 
Multiple experimental models thereof have been developed to investigate whether (and how) biological function and self-organization can emerge from physical principles alone \cite{sanchez_spontaneous_2012, nedelec_self-organization_1997,needleman_active_2017, ruske2021morphology} (e.g., Figure~\ref{fig:schematic}). 
These models offer theoretical insight into out-of-equilibrium phenomena, and could guide the design of new materials that target specific dynamic behaviors or that harvest energy from the system~\cite{ doostmohammadi_stabilization_2016, norton_optimal_2020, guillamat_taming_2017, zhang_spatiotemporal_2021, sciortino_polarity_2023, davis2024active}.

The study of ANs is often qualitative because experimental systems are seldom accessible to in-situ measuring probes. 
Instead, researchers rely on images of fluorescent nematogens. 
These images can be processed into local \cite{decamp_orientational_2015} or global maps \cite{ellis_curvature-induced_2018} of the nematic orientation field. They can also be used to study the characteristics of the motion within the system by comparison to simulations \cite{zhang_spatiotemporal_2021, decamp_orientational_2015}, by the tracking of defects \cite{tan_topological_2019}, or by the estimation of the velocity field via particle-image velocimetry (PIV) \cite{opathalage_self-organized_2019,guillamat_taming_2017, duclos_spontaneous_2018, ilina_cellcell_2020, sciortino_polarity_2023}. 
However, using PIV as a basis for downstream processing requires great care because the estimated velocity is often semiquantitative~\cite{opathalage_self-organized_2019,tan_topological_2019,lemma_statistical_2019, balasubramaniam_investigating_2021, zhang_spatiotemporal_2021}. Moreover, other physical fields like pressure, total force, or boundary stress have not been~measured, albeit some are included in theoretical models.

Quantification is particularly difficult in strongly confined systems. PIV velocity, for example, is not accurate near boundaries nor along directions with poor contrast \cite{opathalage_self-organized_2019, lemma_statistical_2019,wioland_ferromagnetic_2016}. 
Yet, although the underlying mechanisms remain obscure, boundary effects seem to extend well beyond the walls to sway the dynamics of the entire system~\cite{duclos_spontaneous_2018,shankar_topological_2022,vcopar2019topology}: The flow regime correlates with the size of the system~\cite{opathalage_self-organized_2019, duclos_spontaneous_2018, keber_topology_2014, shendruk_dancing_2017}, and the shape of the boundary influences defect dynamics \cite{ellis_curvature-induced_2018, vitelli_defect_2004, pearce_defect_2020, \hardouin}. Any attempt to control ANs \cite{sciortino_polarity_2023,zhang_spatiotemporal_2021} would therefore benefit from probing the physics near boundaries to understand their causal role in the self-organization of these materials. 

\change{Here, we study several long-standing open questions~\cite{shankar_topological_2022} in this context: the causes behind defect nucleation~\cite{opathalage_self-organized_2019,hardouin_active_2022}, the role of pressure in AN systems~\cite{doostmohammadi_active_2018, walton_pressure-driven_2020}, the relative contribution of different forces~\cite{doostmohammadi_active_2018}, and---more generally---how boundaries or material interfaces reshape \mbox{active flows~\cite{guillamat_taming_2017,duclos_spontaneous_2018}.}}

To this end, we propose a framework to measure spatiotemporal maps of velocity, force, pressure, and stresses in AN systems. \change{These include several novel measurements such as the pressure itself, the boundary traction, and the shear stress at the wall.} %
Our force is the total force on the system, rather than the nematic force alone.
In addition, the velocity estimates are able to retrieve information in conditions where PIV fails, especially near the boundary. 
Our framework is based on the formulation of a tailor-made inverse problem that brings together nematohydrodynamics and imaging. The measurements are estimated self-consistently from images alone. We use our novel physical measurements to reveal the causes behind the nucleation of boundary defects. We then show how boundary effects drive the flow of the system and characterize the \mbox{(un-)~steadi}ness of the system dynamics through (local) global pressure gradients. Next, we use the force measurements to estimate the time scales of the systems by comparison with a nematic model for the active~force. These connections can be used to study forces of other origins (e.g., elastic): We finish by studying the drag force exerted on deformable inclusions and using it to estimate the time scale of the fluid-solid interaction.

For concreteness, we consider mixtures of microtubule (MT) filaments and kinesin motors confined to a 2D water-oil interface 
with visible boundaries~\cite{sanchez_spontaneous_2012,tan_topological_2019,opathalage_self-organized_2019,decamp_orientational_2015,zhang_spatiotemporal_2021}. These AN systems can be regarded as minimalistic models for the self-organization of the cytoskeleton, for example during mitosis \cite{nedelec_self-organization_1997}. 
The resulting suspensions form bundles of MTs that extend and fold continuously in a dynamical steady state known as active turbulence (Movie 1) \cite{doostmohammadi_active_2018}. The kinesin motors are the underlying cause of these emergent flows (Figure~\ref{fig:schematic}). \change{While we focus on ``tightly'' confined systems because they remain less understood due to a more severe lack of measurements, we also show that our method applies to bulk nematics and to fluid-solid interactions with deformable inclusions.}

\section{Proposed Framework}
\subsection{Nematohydrodynamic Model} 
In nematohydrodynamics, the Navier-Stokes equations are used to model the motion of fluids driven by active nematogens~\cite{doostmohammadi_active_2018}. This is the case of the MT-kinesin suspension.  
Other than from the pressure $p$, the stresses $\stress$ within the fluid stem from viscosity, nematic activity, and elasticity. They read 
\def\Hfree{\mathrm{\mathbf{H}}}
\begin{align}
\label{eq:stresses}
     \stress_v &= \visc \left( \nabla \v + \nabla^{\top} \v \right), \nonumber \\
     \stress_a &= -\act \nemt, \nonumber \\
     \stress_e &(\nemt, \Hfree),%
\end{align}
where $\Hfree$ is the molecular field, $\nemt$ is the nematic tensor, $\v$ is the velocity, $\mu$ is the viscosity, and $\xi$ the activity. The molecular field accounts for the 
relaxation of the nematic tensor through an elastic free energy that ``penalizes'' inhomogeneities in the orientational order \cite{doostmohammadi_active_2018}. 
The 2D nematic tensor  
\begin{equation} \label{eq:nem_tens}
\nemt = 2 \magn \left( \nemv \otimes \nemv - \mathbb{1}/2 \right)
\end{equation}
provides a macroscopic characterization of the orientational order of the system in terms of the local director field $\nemv(\x,t)$ of the nematogens (Figure~\ref{fig:schematic}), and of the magnitude $\magn(\x,t)$ of this~alignment. %
Since many AN systems are microscopic in scale, inertia is often negligible in comparison to the other sources of stress such as viscosity, $\rey \ll 1$. 

In consideration of these observations, \change{let us} formulate the system $s(\v,p; \f, \g, r)=0$ of \change{Stokes} PDEs, where 
\begin{numcases}{s =}\label{eq:stokes}
	\nabla \cdot \left( \stress_v - p\mathbb{1}  \right) + \f & \text{in $\Omega$} \nonumber \\ 
	\nabla \cdot \v - \source & \text{in $\Omega$} \\
	\v - \g & \text{on $\partial \Omega$} \nonumber
\end{numcases}
\change{with some unknown external force density $\f(\x, t)$ and some unknown values $\g(\x,t)$ for the boundary condition (BCs).} 
Notice that we introduced a source $r(\x,t)$ to \eqref{eq:stokes}; this is to accommodate small 3D flows out of the 2D plane of the system. 

\change{Our only assumption hereafter will be that there is a Stokes fluid \eqref{eq:stokes} underlying the AN system under study.} 
\change{This means that the nematic mesogens act on the fluid entirely through the force~$\f$. In the usual} case where the active motors work at a much faster time scale than that of the elastic relaxation~\cite{doostmohammadi_active_2018, joshi_data-driven_2022}, $\stress_e$ is negligible \change{and} the force~$\f$ stems exclusively from the AN stress $\stress_a$ as per
\begin{equation}\label{eq:force_div}
\f= - \act \nabla \cdot \nemt.    
\end{equation}
\change{More generally, the force would stem from the divergence of the sum $\stress_a+\stress_e$ of active and elastic stresses, instead of $\stress_a$~alone.}

\subsection{Formulation of the Inverse Problem}

\def\mt{m}
We propose to measure all the physical quantities in \eqref{eq:stokes} from images of the AN system acquired with a fluorescence microscope. \change{We will now be formulating an inverse problem to this end.}

Consider that the concentration $\mt(\x,t)$ of MTs must be conserved over time. The differential form of the corresponding continuity equation can be expressed as 
\begin{equation}\label{eq:diff_conservation}
\mathcal{D}_{t,\v}  \{\mt \} = 0
\end{equation}
in terms of the material derivative $\mathcal{D}_{t,\v} \{ \mt \} \define \v \cdot \nabla \mt + \partial_t \mt$. The fluorescence $\fluo (\x,t) \propto m$ emitted by the MTs must, therefore, be conserved too. Since the fluorescence at acquisition is noisy and might exit the imaging plane, we only want to enforce the conservation ``weakly''. Hence, in place of \eqref{eq:diff_conservation}, we formulate the corresponding integral-conservation equation
\begin{equation}\label{eq:conservation}
D(\v; \fluo)= \int_\Omega \mathcal{D}_{t,\v}  \{\fluo \}^2 \approx 0
\end{equation}
over the domain $\Omega \ni \x$ of the AN system, where we use the approximation $\approx$ to enforce the idea of ``weakly''.

Our aim is to measure the physical force $\f$ from the fluorescence $\psi$ in the images. The idea would be to find a velocity $\v$ that (roughly) matches the movement of the fluorescence in the image, $D(\v)\approx 0$, while satisfying the laws of fluid dynamics, $s(\v)=0$. However, $\v$ can only be determined from the PDE model $s=0$ if the force, the values for the BCs, and the source are known. Therefore, the actual problem is to find the unknowns $\f$, $\g$, and $r$ such that the conservation equation is as close to zero as possible ($D$ is minimal) and such that the underlying physics is obeyed. 
This idea can be formalized~as
\begin{equation}\label{eq:opti_before}
\arg \min_{\f,\g,r} D(\v; \fluo)  \ \text{ subject to } \ s(\v,p;\f,\g,r)=0.
\end{equation}
One does not need to optimize for $\v$ or $p$ because they are byproducts of the PDE $s=0$.

Problem \eqref{eq:opti_before} is ``ill-posed'' in the sense that it might have more than one solution. To properly ``pose'' the problem in the face of the multiple unknowns, we incorporate some complementary insight. First, we posit that the MTs cannot cross the boundaries of a confined system. This translates into the \change{weak impermeability} BC $\g \cdot \n \approx 0$, where $\n(\x)$ is the normal vector. In other words, there should be (practically) no flow through the system boundary. (We adapt this to systems with different boundaries in \appx~\ref{sec:freebcs}.) %
And second, sources should be small, $\source \approx 0$, because the movement happens (mostly) within a 2D layer. 

We impose these constraints in a ``weak'' sense by formulating a regularization term $R(\v_0,r,\f) \approx 0$ 
that has to be minimized too (\appx~\ref{sec:app_inv}). 
By incorporating $R$ into \eqref{eq:opti_before}, the final inverse problem becomes 
\begin{equation}\label{eq:opti_problem}
    \arg \min_{\f, \g, r} \left(D + R\right) \ \text{  subject to  } \ s=0.
\end{equation}

\subsection{Resulting Inverse Measurements}
The solution of \eqref{eq:opti_problem} is detailed in \appx~\ref{sec:app_inv}-\ref{sec:app_software}. \change{The input to our method~\eqref{eq:opti_problem} are the fluorescent images.} The \change{(outputted)} physical measurements are not only velocity in the bulk (as in PIV), but also the velocity at the boundaries, as well as the pressure and force field\footnote{\change{Technically, another estimate is the source correction $r^\star(\x,t)$, but we have observed that it is rarely far from zero (strictly) inside the field of view.} }. We refer to them with a star: $\v^\star(\x,t)$, $\g^\star(\x,t)$, $p^\star(\x,t)$, and~$\f^\star(\x,t)$. They comply with the physical equations while accounting for the BCs. 

From these measurements, we derive other quantities such as the pressure gradient $\nabla p^\star (\x,t)$, the shear stress $\stress_v^\star (\x,t)$, the shear stress $\tau_{v, \n_{\perp}}^\star(\x,t)$ that is tangential to the wall, the traction $\trac^\star (\x,t)$ at the boundary, and the drag force $d^\star(t)$ (see \appx~\ref{sec:app_stress} and \suppx~\ref{sec:app_drag} for the definitions). 
To our knowledge, most of these measurements are currently unavailable with the exception of the velocity, especially at or near the boundary. \change{In standard conditions, our measured (total) force $\f^\star(\x,t)$ can be associated to the AN force as per~\eqref{eq:force_div}, which has been estimated before in the literature from estimations of the orientation field~\cite{ellis_curvature-induced_2018}.}

The accuracy of the mathematical framework that we use to solve problem \eqref{eq:opti_problem} was tested in another context using simulations ~\cite{boquet-pujadas_reformulating_2022}. The simulations were performed with $s=0$ being a set of second-order continuum PDEs with the same ellipticity as~\eqref{eq:stokes}, and at 
an equivalent level of ``ill-posendess'' of the inverse problem.

\section{Results}

\begin{figure*}
\centering
    \includegraphics[width=0.95\textwidth]{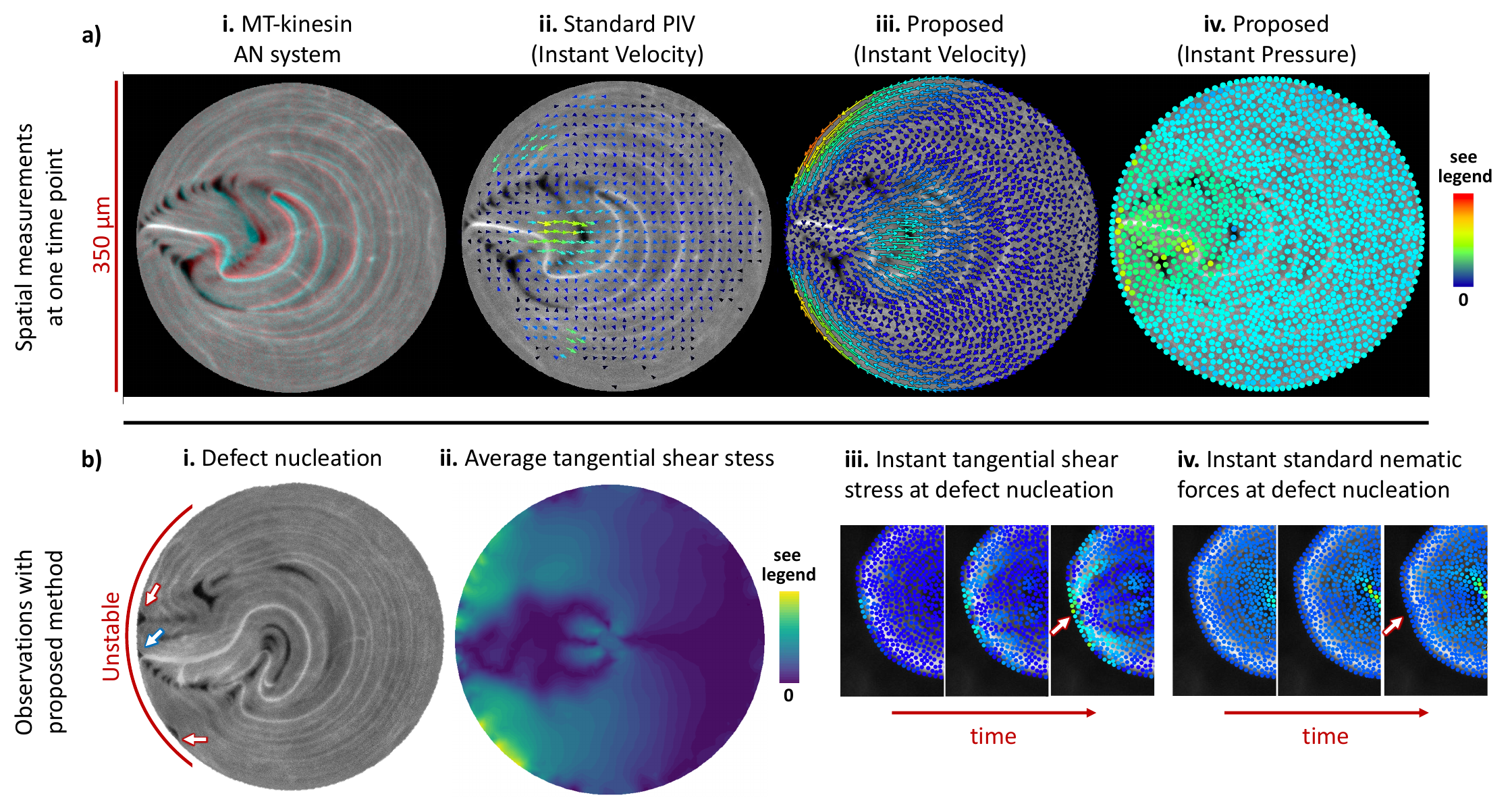}
    \vspace{-5pt}
	\caption{\textbf{New wall and velocity stress measurements near the boundary provide insight into defect nucleation and boundary effects}. \textbf{a.i)} Sum of two consecutive frames (in red and cyan scale, respectively) of a MT-kinesin AN system (first kind, see text) that we imaged with a confocal microscope. Since red and cyan intensities sum to white, ``colorless'' zones correspond to an apparent lack of motion or of spatial contrast between frames (Movie~1), making PIV unreliable. \textbf{a.ii)} Instantaneous velocity computed from the two frames with an iterative PIV method in ImageJ (\appx~\ref{sec:app_software}). \textbf{a.iii-iv)} Instantaneous velocity and pressure computed from the two frames with our inverse method. The colorbar has a range of (\unit{0-61}{\mum / \second}, jet). The pressure (range \unit{0-9.4}{\second^{-1}}, jet) is normalized, $p/\mu$, and up to a constant. \textbf{b.i)} One of the two image frames in (a.i) but with the zones of highest defect nucleation marked in red. The red arrows point at new defects that get absorbed by the main defect (blue arrow), which persists in time. \textbf{b.ii)} The average over time $\bar{\tau}_{v, n_{\perp}}^\star (\x)/\mu$ of the normalized shear stress that is tangential to the boundary (\unit{0-1.0}{\second^{-1}}, viridis). \textbf{b.iii)} Instantaneous $\tau_{v, n_{\perp}}^\star (\x, t)/\mu $ before and during the nucleation (arrow) of a defect (\unit{0-2.0}{\second^{-1}}, jet) (second kind~\cite[Movie~S1]{opathalage_self-organized_2019}). \textbf{b.iv)} For comparison, instantaneous nematic forces at the same three time points computed through the standard method as $\nabla \cdot \nemt^\star$ (see Section~\ref{sec:nematic_force}). They do not seem locally correlated. 
    In all the article, we use the jet (blue-red) and viridis (violet-yellow) colormaps to differentiate between instantaneous and averaged measurements. Differences in the density of arrows or dots reflect magnification. There is another observation about stress around defects  in the \suppx~\ref{sec:app_observations}, Figure~\ref{fig:motile_stress}.
 }
    \label{fig:piv}
\end{figure*}

We first applied our method to take measurements in two kinds of MT-kinesin AN systems confined by a circular boundary. Both have the same constituents, but their assembly protocols differ (\appx~\ref{sec:app_experiments}).  
The first kind (featuring sharper boundaries) results in a transient `steady' state with a permanent defect on the boundary wall that generates a plume of material \cite{hardouin_active_2022} (Movie 1). When other defects nucleate at the boundary, they are quickly assimilated into the main defect. The second kind (featuring more diffuse boundaries) becomes unsteady. It 
generates complex dynamics where spinning defects are periodically disrupted by the nucleation and unbinding of  boundary defects \cite{opathalage_self-organized_2019}. \change{To highlight the robustness of our framework, the video data for the second kind are directly taken from~\cite{opathalage_self-organized_2019}.} The two systems together allow us to compare boundary effects under different regimes of~steadiness (Figures~\ref{fig:piv}-\ref{fig:pressure}). For all these reasons, we usually use global time-averaged measurements to analyze the first kind of systems, and local, instantaneous measurements for the second~kind\footnote{In figures, we portray them using ``viridis'' and ``jet'' colormaps, respectively. The units and the ranges for the colorbars are always in the captions.}.

\subsection{New Velocity Measurements Near System Boundaries}
Resolving the movement near the boundaries of AN systems is challenging because the velocity field aligns with the MTs, a direction in which spatial contrast and texture are poor (Figure~\ref{fig:piv}a.i). 
In general, motion along the direction of MTs is underestimated by standard PIV methods \cite{opathalage_self-organized_2019}. These methods also disregard incompressibility. \change{For example, at the bottom of Figure~\ref{fig:piv}a.ii, PIV measurements point directly opposite to the flow (see Movie~2). These measurements also violate system-wide conservation of mass in the process, which dictates that the flow should \textit{somehow} return to the plume because it is constantly ejecting fluid.}

By contrast, the $\v^\star(\x,t)$ measurements that we obtain from~\eqref{eq:opti_problem} recover a velocity that agrees better with visual assessments of the flow (see Movie~2) and is more consistent with mass conservation (Figure~\ref{fig:piv}a.iii; units and ranges in legends). We also tested the velocity quantitatively by comparing it to manual tracking (\suppx~\ref{sec:accuracy}); and provide further validation via the testing of the force measurements $\f^\star(\x,t)$ in section~\ref{sec:nematic_force}. 

The resulting velocity fields $\v^\star(\x,t)$ are accurate enough to calculate the derivatives necessary for computing the stress everywhere in the system. For the first time, this includes the velocity near the boundary, thereby offering new physical insight.

\subsection{Boundary Stress Induces Defect Nucleation}\label{sec:stressnucleation}
Boundary defects influence the dynamics of confined systems \cite{keber_topology_2014}, but the relevant variables behind their nucleation remain unknown \change{and are the subject of conflicting hypotheses}~\cite{opathalage_self-organized_2019, hardouin_active_2022,hardouin_reconfigurable_2019}, which---so far---have focused exclusively on the role of the system-wide integral of the active-nematic force. 

We found that zones of high \textit{average} stress close to the boundary (Figure~\ref{fig:piv}b.ii) were more prone to nucleating defects (Figure~\ref{fig:piv}b.i, arrows), especially those with high tangential shear stress $\bar{\tau}_{v, n_{\perp}}^\star (\x)$. (This \textit{average} effect can be seen in systems of the first kind because they are steady, Movie~3.) Moreover, among the two main zones of nucleation (arrows), the one below (with the highest stress) was more prolific. This imbalance might explain the tendency of the main wall defect to drift clockwise over time. 

Congruently, the \textit{instantaneous} tangential shear stress $\tau_{v, n_{\perp}}^\star(\x,t)$ was also high before and during nucleation. In Figure~\ref{fig:piv}b.iii (arrow), we show one such example in a system of the second kind for generality; but this applies to those of the first kind, too. 

Like in the buckling of aligned fibers, we hypothesize that bending perturbations might be amplified into boundary defects in these 
conditions of shear. 

\change{Our study of nucleation was possible because our method provides measurements (including shear stress, which is dominant here) on and near the wall. By contrast, \cite{opathalage_self-organized_2019, hardouin_active_2022} formulated two opposing hypotheses based on the correlation with peaks or valleys, respectively, of the integral of the estimated active \textit{force} over the interior domain. For completeness, we studied this force at a local level instead, but we did not find any substantial local correlation preceding defect nucleation (Figure~\ref{fig:piv}b.iv).}

\begin{figure*}
\centering
    \includegraphics[width=0.9\textwidth]{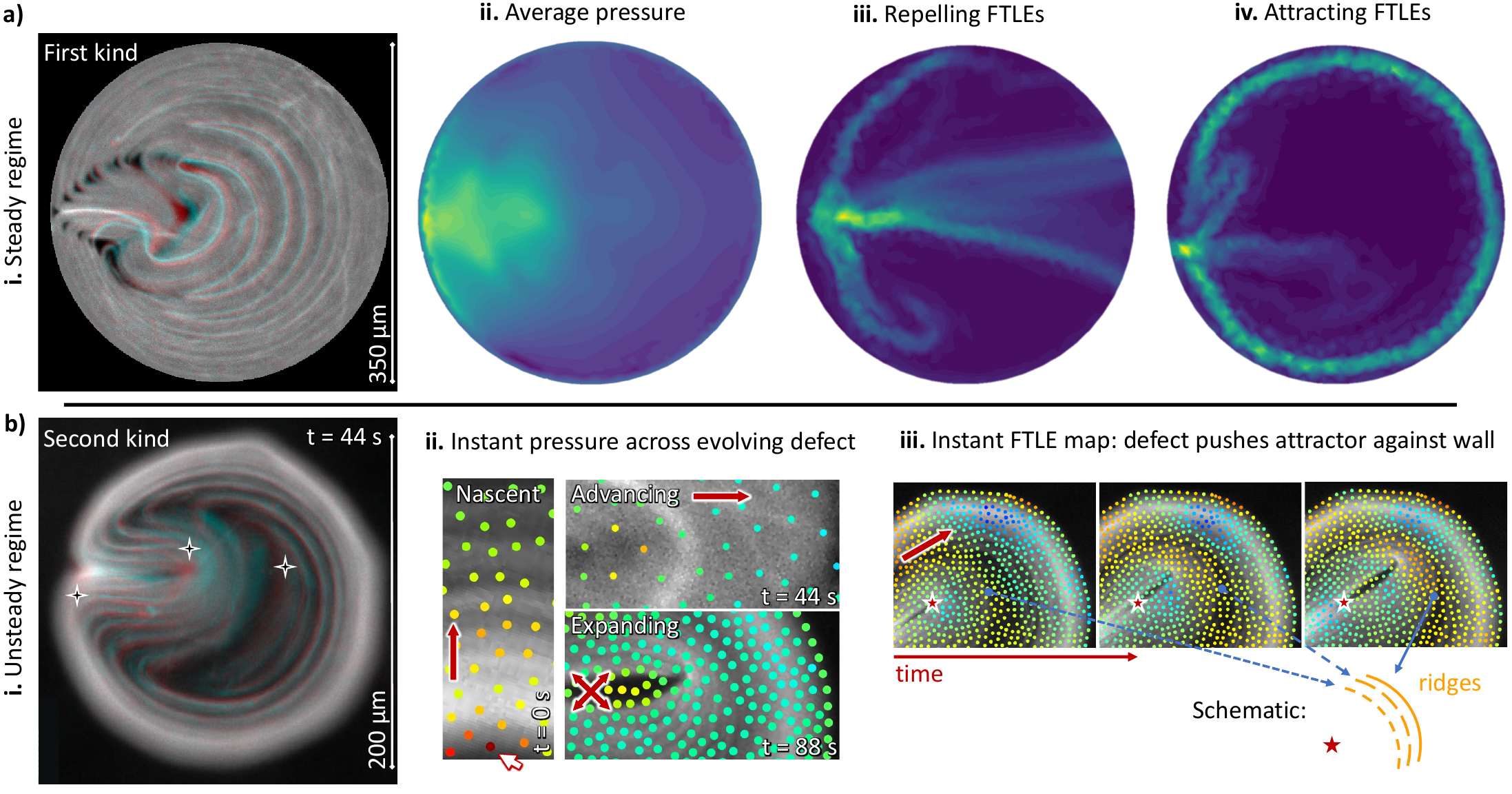}
    \vspace{-5pt}
	\caption{ \textbf{Novel pressure measurements and attractor characterization show how boundaries and defects organize the flow}. \change{\textbf{a)} System of the first kind (Movie~1).} \textbf{a.i)} Sum of two consecutive frames (see Figure \ref{fig:piv}.a.i) of a system of the first kind (see text). These systems have steadier dynamics than those of the second kind. \textbf{a.ii)} Average pressure $\bar{p}^\star (\x) / \mu$ over time in the system (range \unit{0-1.8}{\second^{-1}}, viridis). \textbf{a.iii-iv)} The repelling and attracting FTLEs of the system are plotted as $\surd{\lambda^{\pm}_{\text{max}}(\x)}$ for positive (iii) and negative (iv) time increments (range \unit{0-33}). They barely change over time. \change{\textbf{b)} Second kind (video from~\cite[Movie~S1]{opathalage_self-organized_2019}).} \textbf{b.i)} Sum of two consecutive frames of a system of the second kind (less stable), with periodic nucleation and large-scale flow reorganization. The four-pointed stars mark the location of the defect at the three time points presented in (b.ii). \textbf{b.ii)} Instantaneous, normalized pressure $p^\star (\x,t) / \mu$ (range \unit{0-1.8}{\second^{-1}}, jet) shows gradients across (in the direction of movement of) an evolving defect that nucleated at the boundary. The ``nascent'' defect is rotated \unit{90}{\degree} with respect to (b.i). The white arrow points at the nucleation of a boundary defect (at the wall), the straight arrows indicate the direction of movement, and the crossed arrows indicate outwards expansion (as the defect gets closer to the opposite wall). Smaller densities of dots reflect higher magnifications. \textbf{b.iii)} Instantaneous attracting FTLEs over time plotted as $\ln{\surd{\lambda^{-}_{\text{max}}(\x,t)}}$. %
    The ridges of this FTLE map are unstable manifolds that act as separatrices between different behaviors of flow stretching.  
    For spatial reference, the red star marks the same point in the three time frames. 
 }
 
    \label{fig:pressure}
\end{figure*}
\subsection{Pressure Gradients Characterize Flow Organization}

It is unclear whether and how defects and boundary effects drive the dynamics of ANs \cite{duclos_spontaneous_2018, opathalage_self-organized_2019}. At the same time, the role of pressure in AN systems remains poorly understood and is often neglected~\cite{doostmohammadi_active_2018}, likely for lack of measurements~\cite{walton_pressure-driven_2020}.

We applied our framework to measure the pressure $p^\star(\x,t)$ in systems of the first kind to study steady conditions. Since the pressure did not vary much over time (Figure~\ref{fig:piv}a.iv), we studied its time average $\bar{p}^\star(\x)$. We found that the `steady' boundary defect at the left of the system generates a \change{global} pressure gradient that pushes the material against the opposite wall in the form of a jet or plume (Figure~\ref{fig:pressure}a.ii). The movement is reminiscent of a pair of Rayleigh-B\'enard convection cells. %

\change{For systems of the second kind in the unsteady regime,} pressure gradients \change{worked at} a \change{more} local level instead, accompany\change{ing} defects since their nucleation at the boundary. \change{In particular, we observed pressure gradients across positive defects, but not across negative ones. This was the case for nascent defects at the wall, forward-moving defects leaving the wall, and defects expanding due to interaction with the opposite wall (Figure~\ref{fig:pressure}b.ii).}

\change{The pressure could, therefore, offer an alternative way to characterize the macroscopic dynamics of AN systems, notably during their transition to steady states. In particular, global (local) gradients seems to characterize the organization of steady (unsteady) flow. To further support this hypothesis, in the \suppx~\ref{sec:novisible}, we present a system where global pressure gradients lose their organization as the flow destabilizes.} %

We also studied the effect of pressure in defect nucleation as compared to that of shear discussed in Section~\ref{sec:stressnucleation}. Near the wall opposite to the plume, where pressure and tangential stress are low (Figures~\ref{fig:piv}b.ii,\ref{fig:pressure}a.ii), defects rarely nucleated. Instead, they nucleated preferentially at either side of the main boundary defect, where pressure is low too, but tangential stress is~high.

\subsection{Boundaries Drive the Dynamical System}
\def\t{\Delta t}
\def\fmap{\gamma_{t_0}^{\t}}
\def\Fmap{\Gamma}%
Chaotic flows in ANs have been studied far from the boundary by comparing the topological entropy created by the braiding of defects to the \textit{average} Lyapunov exponent under \textit{ergodic} conditions \cite{tan_topological_2019}. \change{Instead,} to study the \textit{time-varying} unsteady flows \change{that are present} in \change{some of} our systems, we used \textit{finite-time} Lyapunov exponents (FTLEs)~\cite{shadden_definition_2005}. \change{This was only possible due to the accuracy of our velocity fields, especially near the boundary~\cite{boquet_pujadas_variational_2019}.} Repelling (Attracting) FTLEs are a measure of how much flow `particles' that are initially close together separate (converge) as time goes by~\cite{shadden_definition_2005}.  
The result are two maps $\lambda_\text{max}^+(\x, t)$ and $\lambda_\text{max}^-(\x, t)$, respectively, that bring out the topology of the flow and quantify local stretching (\appx~\ref{sec:app_lyapunov}), .

We used FTLEs to complement our pressured-based observations of how defects and boundary effects drive the dynamics of ANs. We first studied the first kind of systems, which display a `steady' regime (Figure~\ref{fig:pressure}a.i). 
Accordingly, the FTLEs did not vary much over time. The repelling FTLEs (\ref{fig:pressure}a.iii) outlined the jet stream formed by the main boundary defect, whereas the circular boundary and the two vortices at either side of the jet were the most attractive zones (Figure~\ref{fig:pressure}a.iv). This reveals that the main wall defect drives the dynamics of the system. It shows that the boundaries recirculate the material into the defect, creating a feedback loop that would not be possible without confinement.

Conversely, the FTLEs in the second kind of (less steady) systems evolved in time. In this case, motile positive defects seemed to organize the flow by deforming the FTLE field. \change{This observation agrees with that in~\cite{serra_defect-mediated_2023}; this article appeared in parallel and provides a rich study of FTLEs in a slow unsteady system, but far from the boundary}. In our case, defects further pushed the attracting regions forward (Figure~\ref{fig:pressure}b.iii) and splattered them against (and around) the boundary. The defects stayed isolated from the wall (in a steady, periodic motion) thereafter.

\subsection{A Comparison between Nematic and Fluid Forces}\label{sec:nematic_force}
\begin{figure}
\centering
\includegraphics[width=0.485\textwidth]{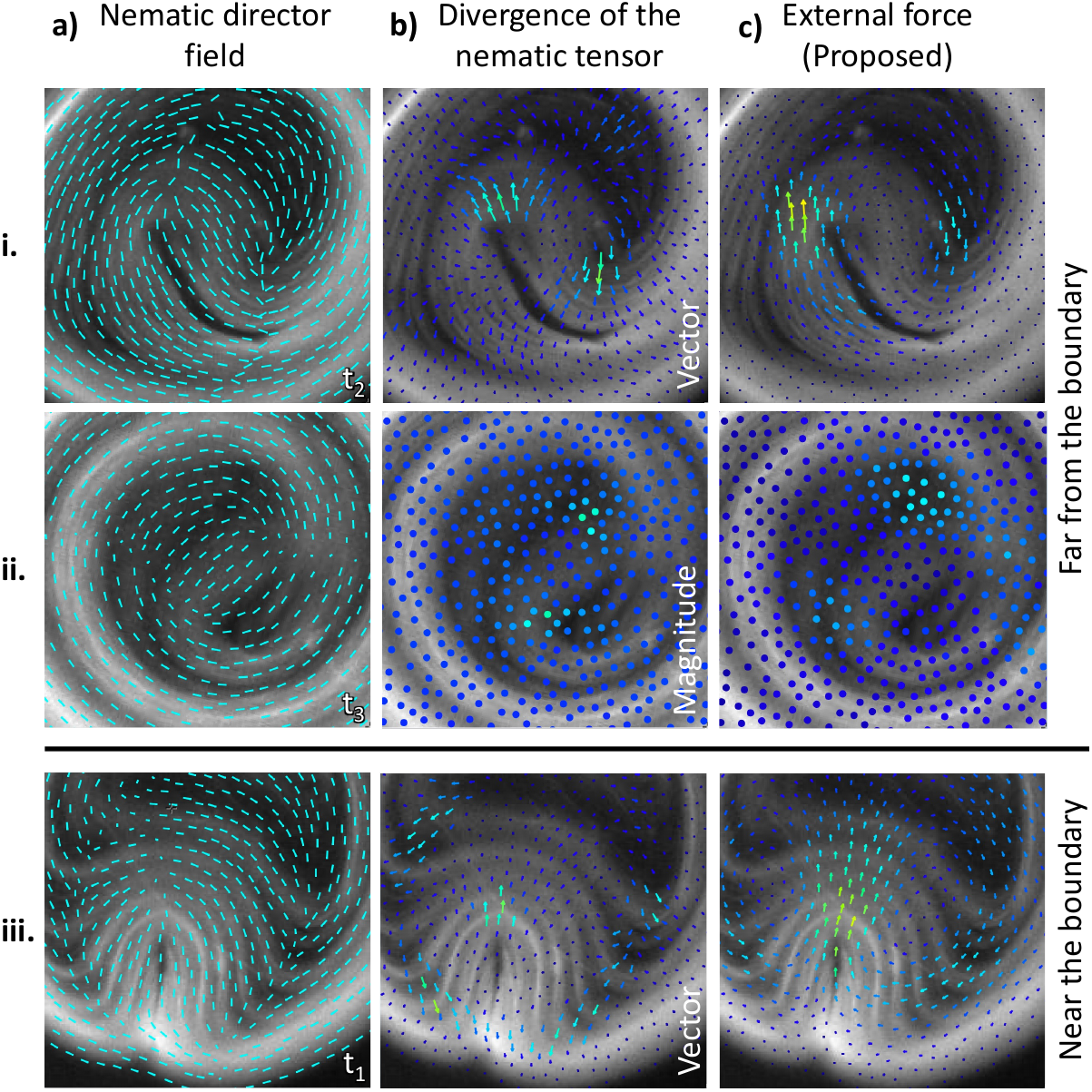}
    \vspace{-5pt}
	\caption{\textbf{New force measurements provide time scale estimates}. \textbf{a)} Local AN director field $\nemv(\x,t)$ obtained from the image-structure tensor at three different times and places. 
    \textbf{b)} Normalized nematic force $-\nabla \cdot \nemt^\star(\x,t)$ computed from the nematic orientation at three different times (i-iii). \textbf{c)} External force $\f^\star(\x,t)/\mu$ obtained with our method (range \unit{0-2\cdot 10^{-8}}{(\meter \cdot \second)^{-1}}) at three different times (i-iii). In comparison to the nematic force, the external force is shifted (our method requires a pair of images) and more diffuse (both use low-pass filters, but differently). \change{The system in this figure is of the second kind (video from~\cite{opathalage_self-organized_2019}).} The time stamps (or the order thereof) are not really relevant here, but they are \unit{0}{\second}, \unit{192}{\second}, and \unit{332}{\second}. The (iib) panel is in magnitude instead of vector form just to offer a different illustration. The width of the images is \unit{140}{\micro \meter}.}
    \label{fig:force}
\end{figure}

Until now, forces in AN systems had only been estimated from the divergence $\nabla \cdot \nemt^\star(\x,t)$ of the nematic tensor \cite{ellis_curvature-induced_2018}. According to equation \eqref{eq:nem_tens}, the nematic tensor $\nemt^\star$ is built from measurements of the molecular director $\nemv^\star(\x,t)$ that are obtained from the same fluorescent images (Figure~\ref{fig:force}a.i-iii, \appx \ref{sec:app_tensor}). While this method is widespread, its accuracy has not been tested independently. 

Therefore, we started by comparing our measurements $\f^\star(\x,t)$ of the force with those resulting from the nematic-tensor method. As per  \eqref{eq:force_div}, we found that the spatio-temporal distributions of $\f^\star$ and $\nabla \cdot \nemt^\star$ were often in agreement, notably around moving defects. (Compare Figure~\ref{fig:force}b.i-ii with c.i-ii.) 
The agreement between the two alternatives could be regarded as a mutual validation. In turn, this also attests the accuracy of the velocity measurements because force estimates are informed by the data less directly (only through the PDE).

\paragraph{The Comparison Yields System Scales:} The ratio $\visc/\act$ between viscous dissipation and activity is an intrinsic time scale of the system that has remained elusive for long \cite{giomi_excitable_2011}. The forces recovered through $\nabla \cdot \nemt^\star$ are defined up to $\act$. Conversely, the measurements $\f^\star$ depend on $\visc$ because they rely on the fluid. Therefore, our comparison between the two methods yields an estimate of $\visc/\act$. \change{For example, we measured a ratio of $\unit{16 \pm 3}{\second}$ for the sequence displayed in Figure \ref{fig:force}.i-ii.} More generally, the systems that we analyzed ranged in the order of $\sim\unit{10-100}{\second}$.
\change{Our measurements of the ratio agree with those of a preliminary version of our method~\cite{boquet_pujadas_variational_2019}, as well as with newer estimates~\cite{ joshi_data-driven_2022}. Moreover, one could choose to study them over space and~time, which could be useful for systems where the activity $\act (\x,t)$ evolves~\cite{shankar2019hydrodynamics}, for example because of gradients in the concentration of ATP.}

\paragraph{The Comparison Could Reveal Other Forces:} Patterning the substrate below ANs allows for some control over them~\cite{thijssen_submersed_2021}. This suggests that active forces cannot be considered in isolation. Our method allows to separate the motion due to active forces from the motion created by stress and pressure that occurs closer to the boundary. In combination with the nematic-tensor method, our force measurements can show how nematic activity transfers to the underlying fluid.

There were a few instances where $\f^\star$ and $\nabla \cdot \nemt^\star$ did not match well. They mostly corresponded to zones near the boundary, where forces may cancel out by symmetry or be balanced by the wall (Figure~\ref{fig:force}.iii). \change{The mismatch might be due to inaccuracies in the director field $\nemv$, which needs to be estimated for the nematic-tensor method. (Find an example of such an error at the bottom left of~\ref{fig:force}.a.iii, where the force (b.iii) is maximal.)} 
\change{An alternative to explain the mismatch could be the presence of other forces; for example of elastic origin, which might be more relevant near the boundaries. In such case---or  in others where the relaxation time is not negligible---the comparison of both methods could yield an estimate of the elastic stress through the simple relation $\nabla \cdot \stress_e(\x,t) =  \f^\star + \act \nabla \cdot \nemt^\star$ by revisiting the reasoning between \eqref{eq:stokes} and \eqref{eq:force_div}.} This also illustrates how our method could be used to test different models for AN forces because, formally, we treat $\f(\x,t)$ as ``any'' external forces in~\eqref{eq:stokes}.

\subsection{Other Effects Captured by the Framework}\label{sec:other_boundary}
Boundary effects are not only present in tightly confined systems. They are also important when the flow interacts with other materials, for example. In the \suppx~\ref{sec:other_systems}, we apply our framework to a system with elastic inclusions and study the fluid-solid interaction through the boundary traction $\trac^\star (\x,t)$ that we compute at the interface. (The inclusions are time-dependent inner boundaries.) In particular, we compute the drag force $d^\star(\x,t)$ to derive the time scale (\unit{22 \pm 12}{\milli \second}) of the interaction, which is the ratio between the viscosity of the fluid and the stiffness of the solid.

Finally, in the \suppx~\ref{sec:novisible}, we study a system far from any visible boundary where the flow becomes progressively disorganized. This allows us to study the change in pressure gradients.

We believe that these two additional systems help further illustrate the usefulness of our framework and the generality of our conclusions.

\section{Discussion}
We have formulated an image-based inverse problem to take new measurements in nematic systems. Namely, stress, traction, pressure, drag, and total force. 
Besides these novel measurements, the estimated velocity is accurate where PIV fails, notably near the~boundary.

Our results show how boundary effects drive confined ANs \change{in both steady and unsteady regimes}. The correlation between shear stress and defect nucleation at the boundary opens a new avenue to design AN systems, and to control their behavior via actuators \cite{ zhang_spatiotemporal_2021, sciortino_polarity_2023}. 
As revealed by the FTLE maps, the ability to control defects is important because they channel boundary effects and play a major role in driving the dynamics of the system. Nematic forces propel positive defects forward or generate pressure gradients to the same effect, albeit only part of these forces translate into effective movement. Step by step, our measurements recount how the flow stretch generated locally by molecular forces interacts with the boundary to drive macroscopic~mixing through topological defects. 

\change{Our study also highlights pressure as an important variable to characterize steady (unsteady) systems, where we have seen that global (local) gradients drive the flow. This is also reflected on how slowly (or rapidly) they reshape the FTLE structures, and is in contrast to the fact that pressure is often deemed a negligible variable in this context~\cite{doostmohammadi_active_2018}, likely because it has remained experimentally inaccessible until now.}

Our force measurements are useful to reveal the time scales of AN systems; for example, via the total force (activity scale), or via the drag force (interaction scale) that can be derived from our traction measurements. Moreover, by comparison to the nematic forces derived from the orientation field, our measurements of the total force could help study forces of different origins (e.g., elastic).

Our framework \change{is able to guess the values of arbitrary BCs} and could be generalized to other systems or branches of physics by tweaking 
the underlying model \eqref{eq:stokes}; for example, to study other active systems such as bacterial biofilms and cell colonies, or other materials entirely~\cite{balasubramaniam_investigating_2021}. \change{Notice that regions where the physical properties of the material might be unknown do not need to be modelled because they can instead be isolated with (potentially time-dependent) inner boundaries (see Section~\ref{sec:other_systems}). This stems from the fact that PDEs are determined by their BCs}. Importantly for experimental studies, 
the measurements can be complemented with 
uncertainty estimates 
by accounting for image noise~\cite{boquet-pujadas_reformulating_2022}.

\section*{Acknowledgments}

We would like to acknowledge the MRSEC Biosynthesis facility at Brandeis University for providing the tubulin. We are grateful to M. Pons, A. LeRoux, and G. Iruela (Universitat de Barcelona) for their assistance in the expression of motor proteins. We are indebted to P. Ellis and A. Fern\'andez-Nieves at Georgia Institute of Technology for the code to detect the director field. We would also like to thank A. Fern\'andez-Nieves for discussions. 

This work was supported, in part, by European Union’s Horizon 2020 Research and Innovation Programme under the Marie Sklodowska-Curie Grant 665807: some parts of the work were based on \cite{boquet_pujadas_variational_2019}. J.I.-M., and F.S. acknowledge funding from MICIU/AEI/10.13039/501100011033 (Grant PID2022-137713NB-C21). The Brandeis University MRSEC Biosynthesis facility is supported by NSF MRSEC 2011846.

All authors discussed the results and approved the manuscript. A.B.-P. conceived the project and developed the framework. J.H. performed the experiments (first kind of MT-kinesin mixtures) and acquired the images under the supervision of J. I.-M. and F.S. A.B.-P. wrote the code and analyzed the image data. J.W. helped in packaging and hosting the software. A.B.-P. prepared the figures and wrote the manuscript. All authors helped revise the manuscript.

\appendix%
\section{Computation and Optimization Methods}
\subsection{Inverse Problem}\label{sec:app_inv}
Consider the concentration $\mt(\x,t): \Omega \times \reals \to \reals_{\geq 0}$ of MTs. As in the main text, \eqref{eq:conservation}, we consider the conservation equation in the weak form
\begin{equation}
D= \norm{\mathcal{D}_{t,\v}  \{\fluo \} }_{\Omega, 2}^2 \approx 0,
\end{equation}
where the subscript refers to the $L^2(\Omega)$ norm. 
For compatibility, we write the regularization term $R$ for the weak constraints listed in the main text as
\begin{equation}\label{eq:regularization}
    R = \boldsymbol{\alpha} \cdot (\norm {\g \cdot \n}_{\partial \Omega, 2}^2, \norm { \source }_{\Omega, 2}^2 , \norm {\f }_{\Omega, 2}^2 ) \approx 0,
\end{equation}
where $\boldsymbol{\alpha} \in \reals^3_{\geq 0}$. Notice that each norm (e.g., $\norm {\f }_{\Omega, 2}^2$) returns a scalar.%

The inverse problem \footnote{As opposed to forward problems (simulations), inverse problems try to guess the cause from the effect using incomplete or noisy data. 
For example, magnetic-resonance imaging reconstructs the concentration of hydrogen from radio-frequency signals, and geologists probe the Earth's mantle with seismographs. Other inverse problems are computed tomography and ultrasound imaging.
Our inverse problem is conceptually similar. The main differences are twofold. First, standard inverse problems fit the data directly with a least-squares error. Instead, here we use a conservation equation for the data term. And second, we use a PDE-constraint instead of the standard system matrix.} is then to solve \eqref{eq:opti_problem} for $\f$, $\g$, $r$ using search spaces that are compatible with~\eqref{eq:stokes} such as $L^2$ and $H^1$. These Sobolev spaces contain all functions for which the integral of the square thereof (respectively,  the derivative of the function, too) is finite.

In the context of \eqref{eq:opti_problem}, the $L^2$ norm of $\f$ in $R$ serves as a low-pass filter against acquisition noise and potential model mismatches. Parameter $\boldsymbol{\alpha}$ controls the strength of this filter and can be chosen optimally according to the standard deviation of the Gaussian noise in the image~\cite{morozov_criteria_1984, boquet-pujadas_reformulating_2022}.

The entire formulation of \eqref{eq:opti_problem} reads similarly to a weak, integral-conservation equation. This is not to the detriment of generality, especially when considering that $s=0$ is usually solved in its energy, integral form (because it provably has the same solutions as its local, differential counterpart). Indeed, many physical laws are derived in their differential (PDE) forms from their original integral (energy) forms: Local conservation translates into global conservation (with an arbitrary domain). Therefore, to obtain a unified framework, we approach the PDEs in their weak form,~too. This is standard when working with the finite-element method. This means that both the PDEs from the model, the conservation, and the constraints are approached in their integral form. All things considered, the final variational problem can be regarded as a task of energy minimization similar to that in the formalism of virtual work~\cite{salencon_virtual_2001}.

The problem establishes a tradeoff between being faithful to the image data and to the constraints in $R$, but the underlying PDEs $s=0$ are enforced tightly. The idea is that information sources are rough, but the physics of the fluid is more trustworthy. In this way, a feedback loop is established between the data and the physics.

The solution of problem \eqref{eq:opti_problem} draws from \cite{boquet-pujadas_reformulating_2022, boquet-pujadas_4d_2022}, and is detailed in \appx~\ref{sec:app_opti}.

We remark that the regularization (or prior) on $\f$ in $R$ can be changed to an $L^1$ term based on total variation (TV)~\cite{chambolle_introduction_2010, boquet-pujadas_reformulating_2022}, which would  promote piecewise constant solutions. The boundary conditions therein can also be adapted to other systems. We also note that we can estimate the error of the measurements obtained from \eqref{eq:opti_problem} by considering the images noise with the Bayesian formalism in~\cite{boquet-pujadas_reformulating_2022}. %
A last comment is that our method can be readily extended to 3D.

\subsection{Boundary Conditions for Non-Confined Systems}\label{sec:freebcs}
\change{When there is no physical boundary visible in the image or when the boundary moves (e.g. an inner solid boundary), the weak ``impermeability'' term is no longer informative for the BC. The border of the image becomes $\partial \Omega$ and the values $\g(\x,t)$ for the BC have no real constraints. In such cases, we propose to replace the term $\norm {\g \cdot \n}_{\partial \Omega, 2}^2$ 
in \eqref{eq:regularization} by a general-purpose weak term 
\begin{align}\label{eq:boundary_gradient}
  &\Big\lVert \nabla (\g)_x -  \big(\nabla {(\g)}_{x} \cdot \n \big)\n \Big\rVert_{\partial \Omega, 2}^2 \nonumber \\ + &\Big\lVert\nabla (\g)_y -  \big(\nabla {(\g)}_{y} \cdot \n \big)\n \Big\rVert_{\partial \Omega, 2}^2,
\end{align}
where $\nabla (\g)_x$ is the vector resulting from taking the gradient of the $x$ component of $\g$. Expression \eqref{eq:boundary_gradient} is essentially the $L^2$ norm of the gradient of the velocity field along the boundary and acts like a low-pass filter in the context of the regularization.}

\subsection{Optimization}\label{sec:app_opti}
We approach the non-convex nature of the optimization problem \eqref{eq:opti_problem} with a multiscale algorithm. At each scale, we use an adjoint strategy to optimize the convex problem that results from linearizing term \eqref{eq:conservation}. This can be done by using the adjoint method~\cite{bereziat_solving_2011,heas_efficient_2016} to compute the gradient and the Hessian to feed a descent algorithm \cite{boquet-pujadas_reformulating_2022}. We also explored another alternative. We realized that the solution to the optimization problem can be formulated as the solution to a system of coupled PDEs: Find $\v,p,\u,q$ such that $s(\v,p)=0$ and $s_\text{adj}(\u,q)=0$. Consider the case where $\boldsymbol{\alpha}_1=0$, for example. Then, the two systems are coupled through the variables $\f=\u/(2\boldsymbol{\alpha}_3)$, $\source = q/(2\boldsymbol{\alpha}_2)$. Here $\u$ and $q$ are the adjoint variables in the adjoint equation system
\begin{numcases}{s_\text{adj} =} \label{eq:adjoint}
	\nabla \cdot \left( \stressadj_v - q\mathbb{1}  \right) - 2 \left( \nabla^\dag \psi \ \v + \partial_t \psi \right) \nabla \psi & \text{in $\Omega$} \nonumber \\ 
	\nabla \cdot \u  & \text{in $\Omega$} \\
	\u & \text{on $\partial \Omega$}, \nonumber
\end{numcases}
where $\stressadj_v = \visc \left( \nabla \u + \nabla^{\top} \u \right)$. Notice that \eqref{eq:adjoint} are also the Stokes equations, but driven by the consistency between the motion and the image data instead of by a physical force. The coupled system behaves similarly to a low-pass filter. The optimization problem results in some optimal parameters $\f^\star$ and $\v_0^\star$. The rest of the measurements---namely $\v^\star$ and $p^\star$--- result from solving $s=0$ thereafter.

In practice, we consider all PDEs in their corresponding weak formulations. We use the finite-element method (FEM) to discretize them. It allows us to tackle arbitrary domains with their corresponding boundary conditions. \change{Note also that the FEM only requires first-order derivatives even if the operators in the equations are second degree. This is made possible by integrating by parts and makes computations more stable.}

\subsection{Implementation}\label{sec:app_software}
The optimization problem was implemented with the help of the FEniCS finite-element library \cite{alnaes_fenics_2015}. 
The input to the resulting software are the image sequence (potentially with metadata) and the boundary of the domain (when it is not rectangular). The boundary is often the result of a segmentation algorithm, for example using ImageJ, Icy, or scikit-image. \change{The boundary does not need to be that of a confined system; since the values for the BCs are guessed, one can also input the rectangular boundary of an image of bulk nematics.} The output are the estimated fields over space and time. We also implemented several of the methods (such as the FTLE) that we developed for post-processing the output measurements.

We compared our method to the iterative PIV plugin in ImageJ (\url{sites.google.com/site/qingzongtseng/piv}), which is in widespread use in the community of ANs.

\subsection{Computation of the Stress}\label{sec:app_stress}
We were especially interested in the deviatoric (shear) stress $\n(\x)  \stress_v^\star(\x)$ across the wall surface with normal $\n$. Its normal and tangential components are $\tau_{v, \n}^\star (\x) = \n(\x)  \stress_v^\star(\x)  \n(\x)$ and $\tau_{v, \n_{\perp}}^\star (\x) = \n(\x)  \stress_v^\star(\x)  \n_{\perp}(\x)$ for $\x \in \partial \Omega$, respectively. For visualization, we extended this measure to $\x \notin \partial \Omega$ by considering the normal $\n(\x)=\n(\x_b)$ at the closest point $\x_b \in \partial \Omega$ to $\x \in \Omega$ that is on the boundary: $\x_b \in \arg \min_{\y \in \partial \Omega} \norm{\y-\x}_2$.
\subsection{Finite-Time Lyapunov Exponents}
\label{sec:app_lyapunov}
Consider the flow map $\fmap(\x_0): \x_0 \mapsto \x(t_0+\t; \x_0,t_0) $, which relates an initial condition $(\x_0, t_0)$ to its position after a time $\t$. It describes the trajectories
within the field defined by $\dot{\x}(t; \x_0,t_0)=\v^\star(\x, t)$. We define the Cauchy-Green tensor of the flow map with respect to the initial condition as %
$\cauchy =\Fmap^\intercal \Fmap$, where $ \Fmap = \nabla_{\x_0} \fmap$. We call $\lambda_{\text{max}}(\x_0, t_0)$ its maximum eigenvalue. The \textit{finite-time} Lyapunov exponent (FTLE) at $\x_0 \in \Omega$ is then $\lambda(\x_0, t_0) = |\t|^{-1}\ln{(\sqrt{\lambda_{\text{max}}(\x_0; t_0, \Delta t)})}$. For $\t >0$ the FTLEs are repelling 
(`stable' manifolds produce ridges), whereas they are attracting for $\t <0$ (`unstable' manifolds do so). The ridges of the FTLE map are, thus, stable or unstable manifolds that act as separatrices between qualitatively different behaviors of particle trajectories. (We write $\lambda^+$ for the forward integration, where $\Delta t >0$, and $\lambda^-$ for the backward one.) FTLEs recover the classical Lyapunov exponents when $\t \to \pm \infty$.

We remark that the computation of FTLEs requires very accurate estimates of the velocity field. And that investigating flow patterns from velocity fields alone is complicated without FTLEs.

So-called Lagrangian coherent structures (LCS) can be extracted by analyzing the spatial distribution of FTLEs. LCSs are the most (locally) attracting or repelling surfaces within the flow. They are the equivalent of the paradigm of stable and unstable manifolds for time-varying unsteady flows, and can help assess chaotic behavior.

\paragraph{Relation to the Nematic Orientation:} A perturbation in the initial condition that has only a small component in the direction of the eigenvalue will quickly align therewith because the unstable direction dominates. 
The direction of nematic orientation is that of maximum stretch because the MT-kinesin mixture is extensile. Therefore, one could contemplate approximating the FTLE from $\sqrt{\lambda_{\text{max}}} \approx | \nemv(\x,t_0+\t) \cdot \Fmap \cdot \nemv(\x_0,t_0) |$. While, in practice, the nematic orientation did seem to align roughly with the ridges of the attractors, the estimates resulting from this approach did not yield accurate estimates in our implementation. 

Notice that the expansive and compressive directions around an hyperbolic point have to mix together in a compact domain, otherwise the expansion is not balanced. Similarly to stirring in fluid dynamics, this happens in the form of braiding movements in active nematics. The braiding of defects has been characterised by its topological entropy and should be lower bounded by the Lyapunov exponent \cite{tan_topological_2019}. This is because the flow in our experiments is 2D and incompressible.

Another possible analysis of the advection field would involve the fitting of a dynamical system into the computed velocity fields. This could be done by representing the fluctuations with respect to the mean with a finite set of modes that maximize kinetic energy, for example via Proper Orthogonal Decomposition (POD).

\subsection{Computation of the Nematic Tensor}
\label{sec:app_tensor}
To measure the nematic vector we used
a widespread method in active nematics \cite[Methods]{ellis_curvature-induced_2018}. The method computes the eigenvectors of a smoothed-out version of the classical image-structure tensor $\nabla \fluo \otimes \nabla \fluo (\x,t)$. One of the two eigenvectors is expected to be orthogonal to the MT bundles, pointing in the direction in which the fluorescence gradient is maximal. The other should align with the MTs, i.e., with the molecular director $\nemv(\x,t)$ (Figure~\ref{fig:force}a). 
The eigenvectors are used to build a measurement $\nemt^\star(\x,t)$ of the nematic tensor as per~\eqref{eq:nem_tens}. 

\subsection{Confined MT-kinesin Mixtures}
\label{sec:app_experiments}
The experimental setup for the first kind of system are the same as in \cite[Methods]{\hardouin}. This includes both the reagents and the microscopy. It is imaged with a confocal microscope. This system has a diameter of \unit{350}{\mum} and is confined by a 3D-printed polymer disk, the walls of which impose a sharp lateral confinement. We refer to \cite[Methods]{opathalage_self-organized_2019} for the setup of the second kind of system. This system is smaller (\unit{200}{\mum}-diameter) and is confined within a microchamber fabricated using photolithography. These pools result in a softer confinement. The system is imaged using a widefield microscope. The videos of the second kind of system were taken from \cite{opathalage_self-organized_2019}.

In principle, the biochemical constituents of the two systems are the same. Yet, the second kind of system is less steady even if the confinement is smaller. We hypothesize this might be due to the confinement method: in the first one, the lateral boundaries are sharp; whereas boundaries impose a diffuse confinement in the second one. %

\change{In all the systems studied in this article, the Reynolds number is small enough that the inertial terms are smaller than machine precision.}

\section{Supporting Experiments}

\subsection{Accuracy of the Velocity}\label{sec:accuracy}
\begin{figure}[!h]
\centering
    \includegraphics[width=0.42\textwidth]{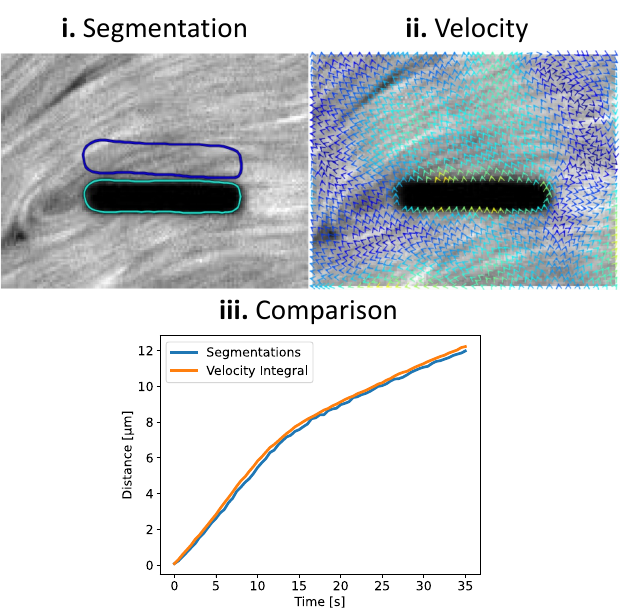}
    \vspace{-5pt}
	\caption{ \change{\textbf{Test of the accuracy of the velocity using a solid inclusion.} \textbf{a)} Segmentations at two different time points. \textbf{b)} Velocity field at a time point (jet color range: \unit{0-2.5}{\micro \meter . \second^{-1}}). \textbf{c)} Displacements computed from the segmentations in (i) compared to the integral of the velocity field estimated by our framework. Both are projected onto the $y$ direction.
 }
 }
    \label{fig:accuracy}
\end{figure}

In addition to the visual tests and to the force comparison, we evaluated the accuracy of our method using the movement of a solid inclusion (see the fluid-solid-interaction experiment presented in the \suppx~\ref{sec:other_systems}). We segmented the inclusion at each time step and computed the position of its center of mass (Figure~\ref{fig:accuracy}i). We compared this to the integral (over time) $\sum_t |\inboundary(t)|^{-1}\int_{\inboundary(t)} \v^\star(\x,t)$ of the velocity at the boundary that we measure with our framework (Figure~\ref{fig:accuracy}ii). The relative error between these two measures was of \unit{4.5\pm 4.4}{\%}. In Figure~\ref{fig:accuracy}iii, we show an example where the final position was \unit{12.0}{\micro \meter} for the segmentation and \unit{12.2}{\micro \meter} for the velocity integral; we consider that the difference is especially low when taking in account that integrating the velocity should accumulate errors over time, and that the two measures might not be exactly the same due to body deformations.

\subsection{System with Elastic Inclusions Relates Fluid Viscosity and Solid Stiffness}\label{sec:other_systems}

\begin{figure}
\centering
    \includegraphics[width=0.43\textwidth]{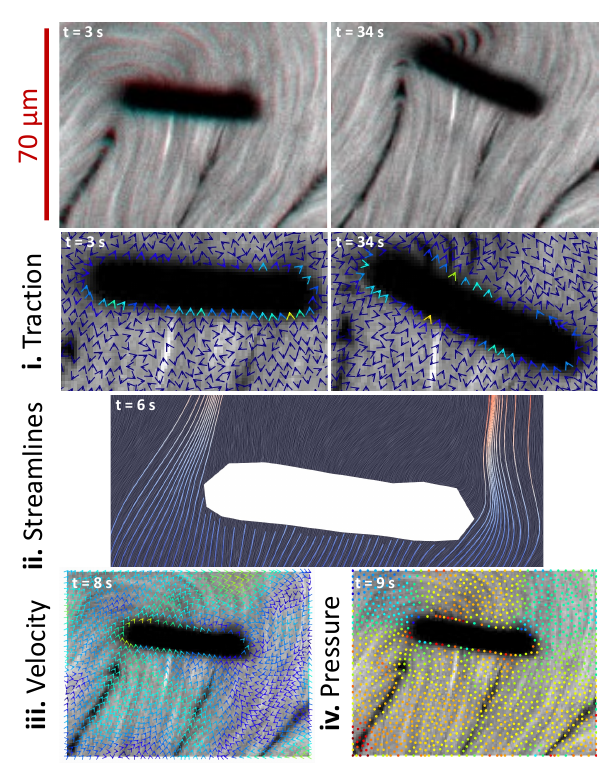}
    \vspace{-5pt}
	\caption{ \change{\textbf{New traction measurements reveal how flows interact with inner solids and provide the viscosity through the drag.} A system with a time-dependent elastic pillar as an inner boundary (Movie~4). Defects push from the bottom by advecting material. A nucleation of a defect at the top left of the pillar happens at \unit{8-9}{\second} (cf. \unit{3}{\second} with \unit{34}{\second}). \textbf{i)} Traction $\trac(\x,t)$ on the boundary before and after the nucleation of a defect (range \unit{0-29}{\second^{-1}}). Due to display limitations arrows appear outside the boundary, where the traction is not defined. However, any arrow that is not on the boundary of the inclusion is exactly $0$ (dark blue). \textbf{ii)} Streamlines of the flow around the pillar colored by integration time from the bottom of the image. Random lines are seeded only at the bottom part of the image to illustrate the shape of the flow. \textbf{iii-iv)} Velocity (\unit{0-3.6}{\micro \meter . \second^{-1}}) and pressure (\unit{0-4.1}{\second^{-1}}).} }
    \label{fig:other_systems}
    \vspace{-10pt}
\end{figure}

\change{
We tested the generality of our framework by studying the interaction of active fluids with solid inclusions. 
To explore the effects of these inner boundaries, we relied on a type of AN preparation inside of which one can photopolymerize compliant, elastic micro-pillars~\cite{velez-ceron_probing_2024}. 
Upon polymerization, these pillars obstruct the flow and are deformed as a result. They constitute a dynamic internal boundary that affects the system (\suppxfig~\ref{fig:other_systems}).} 
\change{We applied our framework to these systems by setting the image frame as the outer boundary and the pillar as a time-dependent inner boundary. In this case we used the alternative BCs described in \appx~\ref{sec:freebcs}---see~\eqref{eq:boundary_gradient}---because the ``weak impermeability'' condition is not adequate.
} 

\change{
The velocity streamlines resulting from our method reveal that the flow wraps around the pillar, albeit not completely as in a typical Stokes-fluid problem because the pillar can move and the flow is active (Figure~\ref{fig:other_systems}ii). We computed the boundary traction $\trac^\star(\x,t)$ (see \suppx~\ref{sec:app_drag}) exerted on the pillar by the AN flow over time and found that it was maximal at the zones where defects ``pushed'' actively against the pillar (Figure~\ref{fig:other_systems}i).}

\change{Similarly to the other systems or setups, the presence of a boundary with such localised stresses made the pillars a place prone to defect nucleation (Figure~\ref{fig:other_systems}iii-iv). These new defects exert a pulling traction on the pillar itself (Figure~\ref{fig:other_systems}i right, at the top of the inclusion).
}

\change{We also computed the (scaled) drag force $d^\star/\mu$ (\unit{11.9 \pm 5.6}{\micro\meter . \second^{-1}}) over time by integrating the traction around the pillar (see \suppx~\ref{sec:app_drag}). 
By comparing the drag force to the stress needed to deform the elastic pillar, we estimated $\mu/E\sim \unit{0.22 \pm 0.12}{\second . \micro \meter}$, where $E$ is the (mesurable) Young's modulus of the material and $\mu$ is the 2D viscosity. This results in a (mean) time scale for the fluid-solid interaction of the system of $\mu_{\text{3D}}/E \sim \unit{22 \pm 12}{\milli \second}$ when considering the viscosity in 3D (see \suppx~\ref{sec:fluidsolidequations}).} 

The inspiration behind this estimation comes from~\cite{velez-ceron_probing_2024}. There, the authors measure the viscosity of the fluid. To this aim, they resort to several approximations due to the lack of pressure and (accurate) velocity measurements. (For example, they use Lamb's equation for cylindrical pillars and use the regional velocity before the pillar is polymerized.) 

If we take the stiffness of the micro-pillar material and use it to estimate the viscosity from our measurements of the time scale of the interaction, we obtain viscosity values that are of the same order as those measured in~\cite{velez-ceron_probing_2024}. Moreover, we observed that the contribution of the pressure to the drag is not always negligible with respect to other sources of stress, contrarily to what is commonly assumed~\cite{doostmohammadi_active_2018,velez-ceron_probing_2024}.

\subsection{Computation of Boundary Traction and Drag}
\label{sec:app_drag}

\change{Let $\indomain(t)$ be a domain (e.g., the moving and deforming pillar) inside the field of view (FOV) such that $\indomain \cup \Omega = \text{FOV}$, where $\Omega$ is the domain of the fluid.}

\change{The traction on the boundary of the pillar, with domain $\inboundary \subset \partial \Omega$, was computed as $\trac^\star(\x,t)= \stress^\star(\x,t) \n (\x,t) $
, $\x \in \inboundary$, where $\stress^\star(\x,t)$ is the sum of stresses. 
To compute the drag forces $\mathbf{d}^\star(\x)$ we projected the traction onto to the general direction of movement $\vcm$ of the center of mass of the pillar (and of the flow), $\mathbf{d}^\star(\x,t) = \trac^\star(\x,t) \cdot \vcm$. (Here, $\n$ and $\vcm$ are unit vectors.) Finally, we compute the total drag as $d^\star(t)=\int_{\inboundary}\mathbf{d}^\star(\x,t)$. Notice that the units of $d^\star/\mu$ are\unit{}{\second^{-1}.\meter}, which means that the drag $d^\star$ has units of force (N) if $\mu$ is taken as a 2D viscosity (with units\unit{}{\pascal.\meter.\second}).}

\change{All these quantities and integrals on the boundary are also computed using the FEM. For this, we take into account that the mesh of the image minus the pillar has to adapt over time as the pillar moves. The remeshing is done with the help of an active-contours segmentation of the pillar over~time.}

\change{Note that there is no need for a model of the material inside the inner boundary $\inboundary$ because the PDE defined on $\Omega$ (which does not contain the domain of the pillar $\indomain$, i.e. inside $\inboundary$) is completely specified by the BCs on $\partial \Omega \supset \inboundary$.}

\change{To establish a link between the viscosity of the AN fluid and the elasticity of the pillar, we had to study the forces involved in the fluid-solid interaction (\suppx~\ref{sec:fluidsolidequations}).}

\subsection{Derivation of the Fluid-Solid Relation for The System with Inclusions}
\label{sec:fluidsolidequations}
In this section, we describe the details of the fluid-solid interaction that we observe in the AN system with the idea of relating $\mu$ to $E$. In what follows, we have tried to detail the logic thoroughly because it could help perform other measurements in the future, but we rely on \eqref{eq:muE} for our computations.

To study the interaction between the active fluid and the elastic pillar, we use that the traction must be balanced at the fluid-solid interface: $\stress \n = - \stress_s \n_s$, where the subscript ``s'' denotes the elastic solid and the lack thereof refers to the fluid.

The equations of the elastic material are
\begin{equation}\label{eq:elasticcontinuum}
    \rho_s \ddot{\v}_s - \nabla \cdot \stress_s(\v_s; \sm) = \mathbf{b},
\end{equation}
\begin{equation}
    \dot{\rho}_s =0,
\end{equation}
with $\rho_s$ and $\sm$ being the $2$D density and shear modulus of the solid. Here, $\mathbf{b}$ is the body force in units of\unit{}{\newton . m^{-2}}. Note that $\v_s$ has units of distance, whereas $\v$ is a velocity. Note also that to reconcile the Lagrangian perspective used for solids with the Eulerian perspective used for liquids, we can use the mapping $\phi: \x_0, t \mapsto \x$ to define the solid displacement as $\v_s(\x_0, t) = \phi(x_0,t)-x_0$ and the fluid displacement as $\v(\x,t)=\v(\phi(x_0,t),t)$, where $x_0$ is the reference point. Then, $\stress_s(\v_s; \sm)= \mathbf{J} \bar{\stress}_s(\boldsymbol{\epsilon}_s; \sm)$, where $\bar{\stress}_s$ is the stress as a function of $2\boldsymbol{\epsilon}_s=\mathbf{J}^\intercal\mathbf{J} - \mathbf{I} $ with $\mathbf{J}=\nabla \phi$. And both frameworks can interact via Piola's transformation at the boundary~\cite{selim_adaptive_2012}. 

If the elastic material was fully contained in the fluid in a ``planar'' way, we would have $\mathbf{b}=0$. However, here the polymerized pillar extends far out in the third dimension and is bent as a result. To compensate for this, we take $\mathbf{b}=F_{\text{bent}}/c_{\text{area}}$, where $c_{\text{area}}$ is the area of the pillar. Then, the force required to deflect a cantilevered beam by $\delta$ is  $F_{\text{bent}}=\delta 3 E I /l^3$, where $E$ is the Young's modulus of the material, which is related to $\sm$. Here, for simplicity,  we supposed\footnote{Otherwise, one can perform the integral or use more specific pre-computed cantilever formulae.} that the force is applied at the tip because the thickness $h$ of the fluid is much smaller than the length $l=\unit{40}{\micro \meter}$ of the pillar~\cite{velez-ceron_probing_2024}, $h<<l$. We can compute the moment of inertia around the neutral axis (and in the direction $\vcm$ of the load) as
\begin{equation}\label{eq:inertia}
I(t)=\int_{\indomain (t)} ((\x-\x_\text{CM}) \cdot \vcm)^2
\end{equation}
using the finite elements that are already in place, where $\x_\text{CM}(t)=|\indomain(t)|^{-1} \int_{\indomain (t)}\x$ is the position of the center of mass of the pillar. Following a similar logic, $\delta(t) =  \vcm \cdot \int_\indomain \v_s(x_0, t)$.

Once the pillar does not accelerate anymore at some time $t=t_f$ (as it stops or changes direction), the momentum term becomes zero because the cantilever force is finally balanced by that of the active fluid. In addition, the relative deformation of the material of the pillar is negligible with respect to the total deflection. Therefore, the integral of the boundary traction must be balanced by the integral of the body force. This boils down to the simpler equation $d(t_f) = 3EI(t_f)\delta(t_f)/l^3$. We can then rearrange to 
\begin{equation}\label{eq:muE}
\frac{\mu}{E}=\frac{3I(t_f)}{l^3} \delta^\star(t_f) \left(\frac{\mu}{d(t_f)} \right),
\end{equation}
where $I(t_f)$, $\delta(t_f)$ and $d(t_f)$ can be computed via integration as detailed in, and after, \eqref{eq:inertia}. For example, we computed $I(t_f)=\unit{3003}{\micro \meter^4}$, $\delta(t_f)=\unit{14.23}{\micro \meter}$, $d^\star(t_f)/\mu = \unit{10.9}{\micro \meter . \second^{-1}}$ for a pillar, and---therefore---$\mu/E= \unit{0.17}{\second . \micro \meter}$. This yields $\mu/E\sim \unit{0.22 \pm 0.12}{\second . \micro \meter}$ over the replicates.

We remark that $\mu$ has units of a 2D problem and $E$ has units of a 3D one. We can express everything in 3D as
\begin{equation}\label{eq:hdivision}
\frac{\mu_{\text{3D}}}{E}=\frac{\mu}{E} \frac{1}{h},
\end{equation}
or---conversely---write it in terms of the Young's modulus in 2D~\cite{lee_measurement_2008}, or in terms of the 2D shear modulus and the 2D Poisson's ratio (e.g., under plane strain conditions~\cite{thorpe_new_1992}). Notice that the resulting division by the thickness $h$ in \eqref{eq:hdivision} is the same that would stem from choosing $\mu$ in 3D from the start and, therefore, having to multiply the drag by $h$ to perform the corresponding 2D (instead of 1D) surface integral. The ratio \eqref{eq:hdivision} has units of time because it describes the time scale of the fluid-solid interaction. Reusing the example above but with the 3D viscosity $\mu_{\text{3D}}$ yields $\mu_{\text{3D}}/E= \unit{0.017}{\second }$ if we take $h=\unit{10}{\micro \meter}$.

The final estimation given in the main text is based on equation \eqref{eq:muE}, but we kept this derivation starting from \eqref{eq:elasticcontinuum} to highlight all the other possible measurements that could be taken.

\begin{figure}[t!]
\centering
    \includegraphics[width=0.42\textwidth]{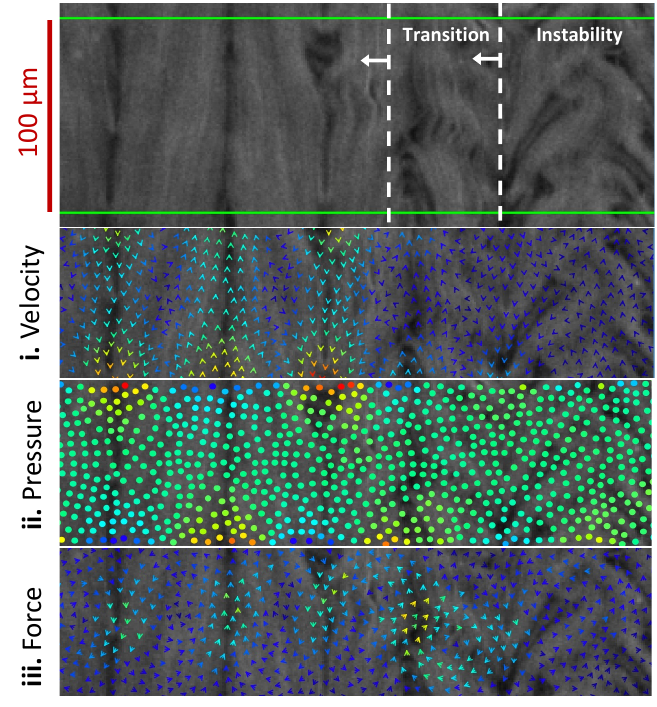}
    \vspace{-5pt}
	\caption{ \change{\textbf{Systems with no visible boundaries} A non-confined system with no visible boundaries that is aligned because of an anisotropy that we introduce to the smectic interface via a magnetic field (Movie~5)~\cite{velez-ceron_probing_2024}. A steady (directed flow) and an unsteady (turbulent flow) zones are present in the same image. The instability is propagating towards the left. Our framework was applied to the green, rectangular region of interest. \textbf{i-iii)} Velocity (\unit{0-24}{\micro \meter . \second^{-1}}), pressure (\unit{0-4.6}{\second^{-1}}), and force fields (\unit{0-2\cdot 10^{-8}}{(\meter \cdot \second)^{-1}}).
 }
 }
    \label{fig:novisibleboundaries}
\end{figure}

\subsection{System with No Visible Boundaries Shows Steady-Unsteady Pressure Transition}
\label{sec:novisible}
\change{We also tested the generality of our framework by studying ANs far from the boundaries. For this, we used yet another MT-kinesin experimental setup: a non-confined, bulk AN system where the flow is aligned in alternating antiparallel directions because of an anisotropy that we introduce to the smectic interface via a magnetic field~\cite{guillamat_control_2016} (\suppxfig~\ref{fig:novisibleboundaries}). Every now and then, the flow destabilizes momentarily. Since there are no physical boundaries in the FOV, we drew an imaginary rectangular boundary and used the same weak prior for the BCs (\appx~\ref{sec:freebcs}) as in the \suppx~\ref{sec:other_systems}. The resulting velocity fields were able to capture the flow despite its spatial heterogeneity, showing that---when aligned---it is driven by a corresponding set of alternating forces (\suppxfig~\ref{fig:novisibleboundaries}.i-iii). The dynamics of the flow are characterized by alternating, vertical pressure gradients (\suppxfig~\ref{fig:novisibleboundaries}.ii, left side of the image), which disappear progressively as the system destabilizes and becomes unsteady and turbulent (right side). This agrees with the conclusions reached in the text about the role of pressure in steady and unsteady systems. We note that, albeit small in magnitude, the pressure is roughly linear in the vertical direction 
and that the gradients extend further if the boundary is chosen (arbitrarily) bigger. }

\subsection{Additional Observations}
\label{sec:app_observations}

In this section, we present two additional observations based on our measurements.

\paragraph{Stress} %

\begin{figure}[h!]
\centering
    \includegraphics[width=0.28\textwidth]{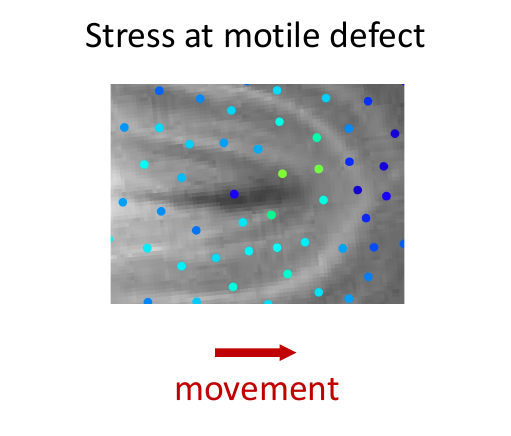}
    \vspace{0pt}
	\caption{ \change{\textbf{The stress reaches extrema at motile defects.} Instantaneous, normalized shear stress $\norm{\stress_v^\star(\x, t)}_\text{F}/\mu$ at a motile defect that has left the boundary (\unit{0-2.0}{\second^{-1}}, jet colormap). Video from~\cite{opathalage_self-organized_2019}.
 }
 }
    \label{fig:motile_stress}
    \vspace{-20pt}
\end{figure}

We observed that motile defects that had left the wall often displayed well-pronounced local minima of stress (blue dot in the middle of Figure~\ref{fig:motile_stress}), %
with the biggest curvature (second order derivative) oriented in the direction of movement. These observations are based on the Frobenius ($\cdot_\text{F}$) norm of the stress, $\norm{\stress_v^\star}_\text{F}$. They are in line with the fact that the MTs bend inwards. %

\paragraph{Force} The forces we measured appeared most prominent where the MT lines fold over themselves, creating (or destroying) `order' as a source of the nematic tensor. %

\section{Description of the Movies}

Find the movies in the Supplemental Material~\cite{supp}.

\noindent \textbf{Movie 1:} Video of a system of the first kind (Figure~\ref{fig:piv}a) as it drifts slightly clockwise. The video lasts \unit{41}{\second}.
\\ \\
\noindent \textbf{Movie 2:} Movie 1 with the velocity field overlaid. Find the scale bar and the range of the colorbar in the caption of Figure~\ref{fig:piv}. The video lasts \unit{41}{\second}.
\\ \\
\noindent \textbf{Movie 3:} Movie 1 with the pressure field overlaid. Find the scale bar and range of the colorbar in the caption of Figure~\ref{fig:piv}. The video lasts \unit{41}{\second}.
\\ \\
\noindent \textbf{Movie 4:} Video of an AN system with an elastic pillar as a solid inclusion. Find the scale bar in Figure~\ref{fig:other_systems}. The video lasts \unit{36}{\second}.
\\ \\
\noindent \textbf{Movie 5:} Video of a non-confined system with no visible boundaries. The system is aligned because of an anisotropy that we introduce to the smectic interface using a magnetic field. Find the scale bar in Figure~\ref{fig:novisibleboundaries}. The video lasts \unit{36}{\second}.

\let\v\oldv
\bibliography{references,sample,references_zotero2,references_prx_supp}

\begin{thebibliography}{60}%
\makeatletter
\providecommand \@ifxundefined [1]{%
 \@ifx{#1\undefined}
}%
\providecommand \@ifnum [1]{%
 \ifnum #1\expandafter \@firstoftwo
 \else \expandafter \@secondoftwo
 \fi
}%
\providecommand \@ifx [1]{%
 \ifx #1\expandafter \@firstoftwo
 \else \expandafter \@secondoftwo
 \fi
}%
\providecommand \natexlab [1]{#1}%
\providecommand \enquote  [1]{``#1''}%
\providecommand \bibnamefont  [1]{#1}%
\providecommand \bibfnamefont [1]{#1}%
\providecommand \citenamefont [1]{#1}%
\providecommand \href@noop [0]{\@secondoftwo}%
\providecommand \href [0]{\begingroup \@sanitize@url \@href}%
\providecommand \@href[1]{\@@startlink{#1}\@@href}%
\providecommand \@@href[1]{\endgroup#1\@@endlink}%
\providecommand \@sanitize@url [0]{\catcode `\\12\catcode `\$12\catcode `\&12\catcode `\#12\catcode `\^12\catcode `\_12\catcode `\%12\relax}%
\providecommand \@@startlink[1]{}%
\providecommand \@@endlink[0]{}%
\providecommand \url  [0]{\begingroup\@sanitize@url \@url }%
\providecommand \@url [1]{\endgroup\@href {#1}{\urlprefix }}%
\providecommand \urlprefix  [0]{URL }%
\providecommand \Eprint [0]{\href }%
\providecommand \doibase [0]{https://doi.org/}%
\providecommand \selectlanguage [0]{\@gobble}%
\providecommand \bibinfo  [0]{\@secondoftwo}%
\providecommand \bibfield  [0]{\@secondoftwo}%
\providecommand \translation [1]{[#1]}%
\providecommand \BibitemOpen [0]{}%
\providecommand \bibitemStop [0]{}%
\providecommand \bibitemNoStop [0]{.\EOS\space}%
\providecommand \EOS [0]{\spacefactor3000\relax}%
\providecommand \BibitemShut  [1]{\csname bibitem#1\endcsname}%
\let\auto@bib@innerbib\@empty
\bibitem [{\citenamefont {Giomi}(2015)}]{giomi_geometry_2015}%
  \BibitemOpen
  \bibfield  {author} {\bibinfo {author} {\bibfnamefont {L.}~\bibnamefont {Giomi}},\ }\bibfield  {title} {\bibinfo {title} {Geometry and {Topology} of {Turbulence} in {Active} {Nematics}},\ }\href {https://doi.org/10.1103/PhysRevX.5.031003} {\bibfield  {journal} {\bibinfo  {journal} {Physical Review X}\ }\textbf {\bibinfo {volume} {5}},\ \bibinfo {pages} {031003} (\bibinfo {year} {2015})}\BibitemShut {NoStop}%
\bibitem [{\citenamefont {Martínez-Prat}\ \emph {et~al.}(2021)\citenamefont {Martínez-Prat}, \citenamefont {Alert}, \citenamefont {Meng}, \citenamefont {Ignés-Mullol}, \citenamefont {Joanny}, \citenamefont {Casademunt}, \citenamefont {Golestanian},\ and\ \citenamefont {Sagués}}]{martinez-prat_scaling_2021}%
  \BibitemOpen
  \bibfield  {author} {\bibinfo {author} {\bibfnamefont {B.}~\bibnamefont {Martínez-Prat}}, \bibinfo {author} {\bibfnamefont {R.}~\bibnamefont {Alert}}, \bibinfo {author} {\bibfnamefont {F.}~\bibnamefont {Meng}}, \bibinfo {author} {\bibfnamefont {J.}~\bibnamefont {Ignés-Mullol}}, \bibinfo {author} {\bibfnamefont {J.-F.}\ \bibnamefont {Joanny}}, \bibinfo {author} {\bibfnamefont {J.}~\bibnamefont {Casademunt}}, \bibinfo {author} {\bibfnamefont {R.}~\bibnamefont {Golestanian}},\ and\ \bibinfo {author} {\bibfnamefont {F.}~\bibnamefont {Sagués}},\ }\bibfield  {title} {\bibinfo {title} {Scaling {Regimes} of {Active} {Turbulence} with {External} {Dissipation}},\ }\href {https://doi.org/10.1103/PhysRevX.11.031065} {\bibfield  {journal} {\bibinfo  {journal} {Physical Review X}\ }\textbf {\bibinfo {volume} {11}},\ \bibinfo {pages} {031065} (\bibinfo {year} {2021})}\BibitemShut {NoStop}%
\bibitem [{\citenamefont {Duclos}\ \emph {et~al.}(2018)\citenamefont {Duclos}, \citenamefont {Blanch-Mercader}, \citenamefont {Yashunsky}, \citenamefont {Salbreux}, \citenamefont {Joanny}, \citenamefont {Prost},\ and\ \citenamefont {Silberzan}}]{duclos_spontaneous_2018}%
  \BibitemOpen
  \bibfield  {author} {\bibinfo {author} {\bibfnamefont {G.}~\bibnamefont {Duclos}}, \bibinfo {author} {\bibfnamefont {C.}~\bibnamefont {Blanch-Mercader}}, \bibinfo {author} {\bibfnamefont {V.}~\bibnamefont {Yashunsky}}, \bibinfo {author} {\bibfnamefont {G.}~\bibnamefont {Salbreux}}, \bibinfo {author} {\bibfnamefont {J.-F.}\ \bibnamefont {Joanny}}, \bibinfo {author} {\bibfnamefont {J.}~\bibnamefont {Prost}},\ and\ \bibinfo {author} {\bibfnamefont {P.}~\bibnamefont {Silberzan}},\ }\bibfield  {title} {{\selectlanguage {en}\bibinfo {title} {Spontaneous shear flow in confined cellular nematics}},\ }\href {https://doi.org/10.1038/s41567-018-0099-7} {\bibfield  {journal} {\bibinfo  {journal} {Nature Physics}\ }\textbf {\bibinfo {volume} {14}},\ \bibinfo {pages} {728} (\bibinfo {year} {2018})}\BibitemShut {NoStop}%
\bibitem [{\citenamefont {Volfson}\ \emph {et~al.}(2008)\citenamefont {Volfson}, \citenamefont {Cookson}, \citenamefont {Hasty},\ and\ \citenamefont {Tsimring}}]{volfson_biomechanical_2008}%
  \BibitemOpen
  \bibfield  {author} {\bibinfo {author} {\bibfnamefont {D.}~\bibnamefont {Volfson}}, \bibinfo {author} {\bibfnamefont {S.}~\bibnamefont {Cookson}}, \bibinfo {author} {\bibfnamefont {J.}~\bibnamefont {Hasty}},\ and\ \bibinfo {author} {\bibfnamefont {L.~S.}\ \bibnamefont {Tsimring}},\ }\bibfield  {title} {\bibinfo {title} {Biomechanical ordering of dense cell populations},\ }\href {https://doi.org/10.1073/pnas.0706805105} {\bibfield  {journal} {\bibinfo  {journal} {Proceedings of the National Academy of Sciences}\ }\textbf {\bibinfo {volume} {105}},\ \bibinfo {pages} {15346} (\bibinfo {year} {2008})}\BibitemShut {NoStop}%
\bibitem [{\citenamefont {Saw}\ \emph {et~al.}(2017)\citenamefont {Saw}, \citenamefont {Doostmohammadi}, \citenamefont {Nier}, \citenamefont {Kocgozlu}, \citenamefont {Thampi}, \citenamefont {Toyama}, \citenamefont {Marcq}, \citenamefont {Lim}, \citenamefont {Yeomans},\ and\ \citenamefont {Ladoux}}]{saw_topological_2017}%
  \BibitemOpen
  \bibfield  {author} {\bibinfo {author} {\bibfnamefont {T.~B.}\ \bibnamefont {Saw}}, \bibinfo {author} {\bibfnamefont {A.}~\bibnamefont {Doostmohammadi}}, \bibinfo {author} {\bibfnamefont {V.}~\bibnamefont {Nier}}, \bibinfo {author} {\bibfnamefont {L.}~\bibnamefont {Kocgozlu}}, \bibinfo {author} {\bibfnamefont {S.}~\bibnamefont {Thampi}}, \bibinfo {author} {\bibfnamefont {Y.}~\bibnamefont {Toyama}}, \bibinfo {author} {\bibfnamefont {P.}~\bibnamefont {Marcq}}, \bibinfo {author} {\bibfnamefont {C.~T.}\ \bibnamefont {Lim}}, \bibinfo {author} {\bibfnamefont {J.~M.}\ \bibnamefont {Yeomans}},\ and\ \bibinfo {author} {\bibfnamefont {B.}~\bibnamefont {Ladoux}},\ }\bibfield  {title} {{\selectlanguage {en}\bibinfo {title} {Topological defects in epithelia govern cell death and extrusion}},\ }\href {https://doi.org/10.1038/nature21718} {\bibfield  {journal} {\bibinfo  {journal} {Nature}\ }\textbf {\bibinfo {volume} {544}},\ \bibinfo {pages} {212} (\bibinfo {year} {2017})}\BibitemShut {NoStop}%
\bibitem [{\citenamefont {Ilina}\ \emph {et~al.}(2020)\citenamefont {Ilina}, \citenamefont {Gritsenko}, \citenamefont {Syga}, \citenamefont {Lippoldt}, \citenamefont {La~Porta}, \citenamefont {Chepizhko}, \citenamefont {Grosser}, \citenamefont {Vullings}, \citenamefont {Bakker}, \citenamefont {Starruß}, \citenamefont {Bult}, \citenamefont {Zapperi}, \citenamefont {Käs}, \citenamefont {Deutsch},\ and\ \citenamefont {Friedl}}]{ilina_cellcell_2020}%
  \BibitemOpen
  \bibfield  {author} {\bibinfo {author} {\bibfnamefont {O.}~\bibnamefont {Ilina}}, \bibinfo {author} {\bibfnamefont {P.~G.}\ \bibnamefont {Gritsenko}}, \bibinfo {author} {\bibfnamefont {S.}~\bibnamefont {Syga}}, \bibinfo {author} {\bibfnamefont {J.}~\bibnamefont {Lippoldt}}, \bibinfo {author} {\bibfnamefont {C.~A.~M.}\ \bibnamefont {La~Porta}}, \bibinfo {author} {\bibfnamefont {O.}~\bibnamefont {Chepizhko}}, \bibinfo {author} {\bibfnamefont {S.}~\bibnamefont {Grosser}}, \bibinfo {author} {\bibfnamefont {M.}~\bibnamefont {Vullings}}, \bibinfo {author} {\bibfnamefont {G.-J.}\ \bibnamefont {Bakker}}, \bibinfo {author} {\bibfnamefont {J.}~\bibnamefont {Starruß}}, \bibinfo {author} {\bibfnamefont {P.}~\bibnamefont {Bult}}, \bibinfo {author} {\bibfnamefont {S.}~\bibnamefont {Zapperi}}, \bibinfo {author} {\bibfnamefont {J.~A.}\ \bibnamefont {Käs}}, \bibinfo {author} {\bibfnamefont {A.}~\bibnamefont {Deutsch}},\ and\ \bibinfo {author} {\bibfnamefont {P.}~\bibnamefont {Friedl}},\ }\bibfield  {title}
  {{\selectlanguage {en}\bibinfo {title} {Cell–cell adhesion and {3D} matrix confinement determine jamming transitions in breast cancer invasion}},\ }\href {https://doi.org/10.1038/s41556-020-0552-6} {\bibfield  {journal} {\bibinfo  {journal} {Nature Cell Biology}\ }\textbf {\bibinfo {volume} {22}},\ \bibinfo {pages} {1103} (\bibinfo {year} {2020})}\BibitemShut {NoStop}%
\bibitem [{\citenamefont {Genkin}\ \emph {et~al.}(2017)\citenamefont {Genkin}, \citenamefont {Sokolov}, \citenamefont {Lavrentovich},\ and\ \citenamefont {Aranson}}]{genkin2017topological}%
  \BibitemOpen
  \bibfield  {author} {\bibinfo {author} {\bibfnamefont {M.~M.}\ \bibnamefont {Genkin}}, \bibinfo {author} {\bibfnamefont {A.}~\bibnamefont {Sokolov}}, \bibinfo {author} {\bibfnamefont {O.~D.}\ \bibnamefont {Lavrentovich}},\ and\ \bibinfo {author} {\bibfnamefont {I.~S.}\ \bibnamefont {Aranson}},\ }\bibfield  {title} {\bibinfo {title} {Topological defects in a living nematic ensnare swimming bacteria},\ }\href@noop {} {\bibfield  {journal} {\bibinfo  {journal} {Physical Review X}\ }\textbf {\bibinfo {volume} {7}},\ \bibinfo {pages} {011029} (\bibinfo {year} {2017})}\BibitemShut {NoStop}%
\bibitem [{\citenamefont {Sanchez}\ \emph {et~al.}(2012)\citenamefont {Sanchez}, \citenamefont {Chen}, \citenamefont {DeCamp}, \citenamefont {Heymann},\ and\ \citenamefont {Dogic}}]{sanchez_spontaneous_2012}%
  \BibitemOpen
  \bibfield  {author} {\bibinfo {author} {\bibfnamefont {T.}~\bibnamefont {Sanchez}}, \bibinfo {author} {\bibfnamefont {D.~T.~N.}\ \bibnamefont {Chen}}, \bibinfo {author} {\bibfnamefont {S.~J.}\ \bibnamefont {DeCamp}}, \bibinfo {author} {\bibfnamefont {M.}~\bibnamefont {Heymann}},\ and\ \bibinfo {author} {\bibfnamefont {Z.}~\bibnamefont {Dogic}},\ }\bibfield  {title} {{\selectlanguage {en}\bibinfo {title} {Spontaneous motion in hierarchically assembled active matter}},\ }\href {https://doi.org/10.1038/nature11591} {\bibfield  {journal} {\bibinfo  {journal} {Nature}\ }\textbf {\bibinfo {volume} {491}},\ \bibinfo {pages} {431} (\bibinfo {year} {2012})}\BibitemShut {NoStop}%
\bibitem [{\citenamefont {Nédélec}\ \emph {et~al.}(1997)\citenamefont {Nédélec}, \citenamefont {Surrey}, \citenamefont {Maggs},\ and\ \citenamefont {Leibler}}]{nedelec_self-organization_1997}%
  \BibitemOpen
  \bibfield  {author} {\bibinfo {author} {\bibfnamefont {F.~J.}\ \bibnamefont {Nédélec}}, \bibinfo {author} {\bibfnamefont {T.}~\bibnamefont {Surrey}}, \bibinfo {author} {\bibfnamefont {A.~C.}\ \bibnamefont {Maggs}},\ and\ \bibinfo {author} {\bibfnamefont {S.}~\bibnamefont {Leibler}},\ }\bibfield  {title} {{\selectlanguage {en}\bibinfo {title} {Self-organization of microtubules and motors}},\ }\href {https://doi.org/10.1038/38532} {\bibfield  {journal} {\bibinfo  {journal} {Nature}\ }\textbf {\bibinfo {volume} {389}},\ \bibinfo {pages} {305} (\bibinfo {year} {1997})}\BibitemShut {NoStop}%
\bibitem [{\citenamefont {Needleman}\ and\ \citenamefont {Dogic}(2017)}]{needleman_active_2017}%
  \BibitemOpen
  \bibfield  {author} {\bibinfo {author} {\bibfnamefont {D.}~\bibnamefont {Needleman}}\ and\ \bibinfo {author} {\bibfnamefont {Z.}~\bibnamefont {Dogic}},\ }\bibfield  {title} {{\selectlanguage {en}\bibinfo {title} {Active matter at the interface between materials science and cell biology}},\ }\href {https://doi.org/10.1038/natrevmats.2017.48} {\bibfield  {journal} {\bibinfo  {journal} {Nature Reviews Materials}\ }\textbf {\bibinfo {volume} {2}},\ \bibinfo {pages} {1} (\bibinfo {year} {2017})}\BibitemShut {NoStop}%
\bibitem [{\citenamefont {Ruske}\ and\ \citenamefont {Yeomans}(2021)}]{ruske2021morphology}%
  \BibitemOpen
  \bibfield  {author} {\bibinfo {author} {\bibfnamefont {L.~J.}\ \bibnamefont {Ruske}}\ and\ \bibinfo {author} {\bibfnamefont {J.~M.}\ \bibnamefont {Yeomans}},\ }\bibfield  {title} {\bibinfo {title} {Morphology of active deformable 3d droplets},\ }\href@noop {} {\bibfield  {journal} {\bibinfo  {journal} {Physical Review X}\ }\textbf {\bibinfo {volume} {11}},\ \bibinfo {pages} {021001} (\bibinfo {year} {2021})}\BibitemShut {NoStop}%
\bibitem [{\citenamefont {Doostmohammadi}\ \emph {et~al.}(2016)\citenamefont {Doostmohammadi}, \citenamefont {Adamer}, \citenamefont {Thampi},\ and\ \citenamefont {Yeomans}}]{doostmohammadi_stabilization_2016}%
  \BibitemOpen
  \bibfield  {author} {\bibinfo {author} {\bibfnamefont {A.}~\bibnamefont {Doostmohammadi}}, \bibinfo {author} {\bibfnamefont {M.~F.}\ \bibnamefont {Adamer}}, \bibinfo {author} {\bibfnamefont {S.~P.}\ \bibnamefont {Thampi}},\ and\ \bibinfo {author} {\bibfnamefont {J.~M.}\ \bibnamefont {Yeomans}},\ }\bibfield  {title} {{\selectlanguage {en}\bibinfo {title} {Stabilization of active matter by flow-vortex lattices and defect ordering}},\ }\href {https://doi.org/10.1038/ncomms10557} {\bibfield  {journal} {\bibinfo  {journal} {Nature Communications}\ }\textbf {\bibinfo {volume} {7}},\ \bibinfo {pages} {10557} (\bibinfo {year} {2016})}\BibitemShut {NoStop}%
\bibitem [{\citenamefont {Norton}\ \emph {et~al.}(2020)\citenamefont {Norton}, \citenamefont {Grover}, \citenamefont {Hagan},\ and\ \citenamefont {Fraden}}]{norton_optimal_2020}%
  \BibitemOpen
  \bibfield  {author} {\bibinfo {author} {\bibfnamefont {M.~M.}\ \bibnamefont {Norton}}, \bibinfo {author} {\bibfnamefont {P.}~\bibnamefont {Grover}}, \bibinfo {author} {\bibfnamefont {M.~F.}\ \bibnamefont {Hagan}},\ and\ \bibinfo {author} {\bibfnamefont {S.}~\bibnamefont {Fraden}},\ }\bibfield  {title} {\bibinfo {title} {Optimal {Control} of {Active} {Nematics}},\ }\href {https://doi.org/10.1103/PhysRevLett.125.178005} {\bibfield  {journal} {\bibinfo  {journal} {Physical Review Letters}\ }\textbf {\bibinfo {volume} {125}},\ \bibinfo {pages} {178005} (\bibinfo {year} {2020})}\BibitemShut {NoStop}%
\bibitem [{\citenamefont {Guillamat}\ \emph {et~al.}(2017)\citenamefont {Guillamat}, \citenamefont {Ignés-Mullol},\ and\ \citenamefont {Sagués}}]{guillamat_taming_2017}%
  \BibitemOpen
  \bibfield  {author} {\bibinfo {author} {\bibfnamefont {P.}~\bibnamefont {Guillamat}}, \bibinfo {author} {\bibfnamefont {J.}~\bibnamefont {Ignés-Mullol}},\ and\ \bibinfo {author} {\bibfnamefont {F.}~\bibnamefont {Sagués}},\ }\bibfield  {title} {{\selectlanguage {en}\bibinfo {title} {Taming active turbulence with patterned soft interfaces}},\ }\href {https://doi.org/10.1038/s41467-017-00617-1} {\bibfield  {journal} {\bibinfo  {journal} {Nature Communications}\ }\textbf {\bibinfo {volume} {8}},\ \bibinfo {pages} {564} (\bibinfo {year} {2017})}\BibitemShut {NoStop}%
\bibitem [{\citenamefont {Zhang}\ \emph {et~al.}(2021)\citenamefont {Zhang}, \citenamefont {Redford}, \citenamefont {Ruijgrok}, \citenamefont {Kumar}, \citenamefont {Mozaffari}, \citenamefont {Zemsky}, \citenamefont {Dinner}, \citenamefont {Vitelli}, \citenamefont {Bryant}, \citenamefont {Gardel},\ and\ \citenamefont {de~Pablo}}]{zhang_spatiotemporal_2021}%
  \BibitemOpen
  \bibfield  {author} {\bibinfo {author} {\bibfnamefont {R.}~\bibnamefont {Zhang}}, \bibinfo {author} {\bibfnamefont {S.~A.}\ \bibnamefont {Redford}}, \bibinfo {author} {\bibfnamefont {P.~V.}\ \bibnamefont {Ruijgrok}}, \bibinfo {author} {\bibfnamefont {N.}~\bibnamefont {Kumar}}, \bibinfo {author} {\bibfnamefont {A.}~\bibnamefont {Mozaffari}}, \bibinfo {author} {\bibfnamefont {S.}~\bibnamefont {Zemsky}}, \bibinfo {author} {\bibfnamefont {A.~R.}\ \bibnamefont {Dinner}}, \bibinfo {author} {\bibfnamefont {V.}~\bibnamefont {Vitelli}}, \bibinfo {author} {\bibfnamefont {Z.}~\bibnamefont {Bryant}}, \bibinfo {author} {\bibfnamefont {M.~L.}\ \bibnamefont {Gardel}},\ and\ \bibinfo {author} {\bibfnamefont {J.~J.}\ \bibnamefont {de~Pablo}},\ }\bibfield  {title} {{\selectlanguage {en}\bibinfo {title} {Spatiotemporal control of liquid crystal structure and dynamics through activity patterning}},\ }\href {https://doi.org/10.1038/s41563-020-00901-4} {\bibfield  {journal} {\bibinfo  {journal} {Nature Materials}\ }\textbf
  {\bibinfo {volume} {20}},\ \bibinfo {pages} {875} (\bibinfo {year} {2021})}\BibitemShut {NoStop}%
\bibitem [{\citenamefont {Sciortino}\ \emph {et~al.}(2023)\citenamefont {Sciortino}, \citenamefont {Neumann}, \citenamefont {Krüger}, \citenamefont {Maryshev}, \citenamefont {Teshima}, \citenamefont {Wolfrum}, \citenamefont {Frey},\ and\ \citenamefont {Bausch}}]{sciortino_polarity_2023}%
  \BibitemOpen
  \bibfield  {author} {\bibinfo {author} {\bibfnamefont {A.}~\bibnamefont {Sciortino}}, \bibinfo {author} {\bibfnamefont {L.~J.}\ \bibnamefont {Neumann}}, \bibinfo {author} {\bibfnamefont {T.}~\bibnamefont {Krüger}}, \bibinfo {author} {\bibfnamefont {I.}~\bibnamefont {Maryshev}}, \bibinfo {author} {\bibfnamefont {T.~F.}\ \bibnamefont {Teshima}}, \bibinfo {author} {\bibfnamefont {B.}~\bibnamefont {Wolfrum}}, \bibinfo {author} {\bibfnamefont {E.}~\bibnamefont {Frey}},\ and\ \bibinfo {author} {\bibfnamefont {A.~R.}\ \bibnamefont {Bausch}},\ }\bibfield  {title} {{\selectlanguage {en}\bibinfo {title} {Polarity and chirality control of an active fluid by passive nematic defects}},\ }\href {https://doi.org/10.1038/s41563-022-01432-w} {\bibfield  {journal} {\bibinfo  {journal} {Nature Materials}\ }\textbf {\bibinfo {volume} {22}},\ \bibinfo {pages} {260} (\bibinfo {year} {2023})}\BibitemShut {NoStop}%
\bibitem [{\citenamefont {Davis}\ \emph {et~al.}(2024)\citenamefont {Davis}, \citenamefont {Proesmans},\ and\ \citenamefont {Fodor}}]{davis2024active}%
  \BibitemOpen
  \bibfield  {author} {\bibinfo {author} {\bibfnamefont {L.~K.}\ \bibnamefont {Davis}}, \bibinfo {author} {\bibfnamefont {K.}~\bibnamefont {Proesmans}},\ and\ \bibinfo {author} {\bibfnamefont {{\'E}.}~\bibnamefont {Fodor}},\ }\bibfield  {title} {\bibinfo {title} {Active matter under control: Insights from response theory},\ }\href@noop {} {\bibfield  {journal} {\bibinfo  {journal} {Physical Review X}\ }\textbf {\bibinfo {volume} {14}},\ \bibinfo {pages} {011012} (\bibinfo {year} {2024})}\BibitemShut {NoStop}%
\bibitem [{\citenamefont {DeCamp}\ \emph {et~al.}(2015)\citenamefont {DeCamp}, \citenamefont {Redner}, \citenamefont {Baskaran}, \citenamefont {Hagan},\ and\ \citenamefont {Dogic}}]{decamp_orientational_2015}%
  \BibitemOpen
  \bibfield  {author} {\bibinfo {author} {\bibfnamefont {S.~J.}\ \bibnamefont {DeCamp}}, \bibinfo {author} {\bibfnamefont {G.~S.}\ \bibnamefont {Redner}}, \bibinfo {author} {\bibfnamefont {A.}~\bibnamefont {Baskaran}}, \bibinfo {author} {\bibfnamefont {M.~F.}\ \bibnamefont {Hagan}},\ and\ \bibinfo {author} {\bibfnamefont {Z.}~\bibnamefont {Dogic}},\ }\bibfield  {title} {{\selectlanguage {en}\bibinfo {title} {Orientational order of motile defects in active nematics}},\ }\href {https://doi.org/10.1038/nmat4387} {\bibfield  {journal} {\bibinfo  {journal} {Nature Materials}\ }\textbf {\bibinfo {volume} {14}},\ \bibinfo {pages} {1110} (\bibinfo {year} {2015})}\BibitemShut {NoStop}%
\bibitem [{\citenamefont {Ellis}\ \emph {et~al.}(2018)\citenamefont {Ellis}, \citenamefont {Pearce}, \citenamefont {Chang}, \citenamefont {Goldsztein}, \citenamefont {Giomi},\ and\ \citenamefont {Fernandez-Nieves}}]{ellis_curvature-induced_2018}%
  \BibitemOpen
  \bibfield  {author} {\bibinfo {author} {\bibfnamefont {P.~W.}\ \bibnamefont {Ellis}}, \bibinfo {author} {\bibfnamefont {D.~J.~G.}\ \bibnamefont {Pearce}}, \bibinfo {author} {\bibfnamefont {Y.-W.}\ \bibnamefont {Chang}}, \bibinfo {author} {\bibfnamefont {G.}~\bibnamefont {Goldsztein}}, \bibinfo {author} {\bibfnamefont {L.}~\bibnamefont {Giomi}},\ and\ \bibinfo {author} {\bibfnamefont {A.}~\bibnamefont {Fernandez-Nieves}},\ }\bibfield  {title} {{\selectlanguage {en}\bibinfo {title} {Curvature-induced defect unbinding and dynamics in active nematic toroids}},\ }\href {https://doi.org/10.1038/nphys4276} {\bibfield  {journal} {\bibinfo  {journal} {Nature Physics}\ }\textbf {\bibinfo {volume} {14}},\ \bibinfo {pages} {85} (\bibinfo {year} {2018})}\BibitemShut {NoStop}%
\bibitem [{\citenamefont {Tan}\ \emph {et~al.}(2019)\citenamefont {Tan}, \citenamefont {Roberts}, \citenamefont {Smith}, \citenamefont {Olvera}, \citenamefont {Arteaga}, \citenamefont {Fortini}, \citenamefont {Mitchell},\ and\ \citenamefont {Hirst}}]{tan_topological_2019}%
  \BibitemOpen
  \bibfield  {author} {\bibinfo {author} {\bibfnamefont {A.~J.}\ \bibnamefont {Tan}}, \bibinfo {author} {\bibfnamefont {E.}~\bibnamefont {Roberts}}, \bibinfo {author} {\bibfnamefont {S.~A.}\ \bibnamefont {Smith}}, \bibinfo {author} {\bibfnamefont {U.~A.}\ \bibnamefont {Olvera}}, \bibinfo {author} {\bibfnamefont {J.}~\bibnamefont {Arteaga}}, \bibinfo {author} {\bibfnamefont {S.}~\bibnamefont {Fortini}}, \bibinfo {author} {\bibfnamefont {K.~A.}\ \bibnamefont {Mitchell}},\ and\ \bibinfo {author} {\bibfnamefont {L.~S.}\ \bibnamefont {Hirst}},\ }\bibfield  {title} {{\selectlanguage {en}\bibinfo {title} {Topological chaos in active nematics}},\ }\href {https://doi.org/10.1038/s41567-019-0600-y} {\bibfield  {journal} {\bibinfo  {journal} {Nature Physics}\ }\textbf {\bibinfo {volume} {15}},\ \bibinfo {pages} {1033} (\bibinfo {year} {2019})}\BibitemShut {NoStop}%
\bibitem [{\citenamefont {Opathalage}\ \emph {et~al.}(2019)\citenamefont {Opathalage}, \citenamefont {Norton}, \citenamefont {Juniper}, \citenamefont {Langeslay}, \citenamefont {Aghvami}, \citenamefont {Fraden},\ and\ \citenamefont {Dogic}}]{opathalage_self-organized_2019}%
  \BibitemOpen
  \bibfield  {author} {\bibinfo {author} {\bibfnamefont {A.}~\bibnamefont {Opathalage}}, \bibinfo {author} {\bibfnamefont {M.~M.}\ \bibnamefont {Norton}}, \bibinfo {author} {\bibfnamefont {M.~P.~N.}\ \bibnamefont {Juniper}}, \bibinfo {author} {\bibfnamefont {B.}~\bibnamefont {Langeslay}}, \bibinfo {author} {\bibfnamefont {S.~A.}\ \bibnamefont {Aghvami}}, \bibinfo {author} {\bibfnamefont {S.}~\bibnamefont {Fraden}},\ and\ \bibinfo {author} {\bibfnamefont {Z.}~\bibnamefont {Dogic}},\ }\bibfield  {title} {\bibinfo {title} {Self-organized dynamics and the transition to turbulence of confined active nematics},\ }\href {https://doi.org/10.1073/pnas.1816733116} {\bibfield  {journal} {\bibinfo  {journal} {Proceedings of the National Academy of Sciences}\ }\textbf {\bibinfo {volume} {116}},\ \bibinfo {pages} {4788} (\bibinfo {year} {2019})}\BibitemShut {NoStop}%
\bibitem [{\citenamefont {Lemma}\ \emph {et~al.}(2019)\citenamefont {Lemma}, \citenamefont {DeCamp}, \citenamefont {You}, \citenamefont {Giomi},\ and\ \citenamefont {Dogic}}]{lemma_statistical_2019}%
  \BibitemOpen
  \bibfield  {author} {\bibinfo {author} {\bibfnamefont {L.~M.}\ \bibnamefont {Lemma}}, \bibinfo {author} {\bibfnamefont {S.~J.}\ \bibnamefont {DeCamp}}, \bibinfo {author} {\bibfnamefont {Z.}~\bibnamefont {You}}, \bibinfo {author} {\bibfnamefont {L.}~\bibnamefont {Giomi}},\ and\ \bibinfo {author} {\bibfnamefont {Z.}~\bibnamefont {Dogic}},\ }\bibfield  {title} {{\selectlanguage {en}\bibinfo {title} {Statistical properties of autonomous flows in {2D} active nematics}},\ }\href {https://doi.org/10.1039/C8SM01877D} {\bibfield  {journal} {\bibinfo  {journal} {Soft Matter}\ }\textbf {\bibinfo {volume} {15}},\ \bibinfo {pages} {3264} (\bibinfo {year} {2019})}\BibitemShut {NoStop}%
\bibitem [{\citenamefont {Balasubramaniam}\ \emph {et~al.}(2021)\citenamefont {Balasubramaniam}, \citenamefont {Doostmohammadi}, \citenamefont {Saw}, \citenamefont {Narayana}, \citenamefont {Mueller}, \citenamefont {Dang}, \citenamefont {Thomas}, \citenamefont {Gupta}, \citenamefont {Sonam}, \citenamefont {Yap}, \citenamefont {Toyama}, \citenamefont {Mège}, \citenamefont {Yeomans},\ and\ \citenamefont {Ladoux}}]{balasubramaniam_investigating_2021}%
  \BibitemOpen
  \bibfield  {author} {\bibinfo {author} {\bibfnamefont {L.}~\bibnamefont {Balasubramaniam}}, \bibinfo {author} {\bibfnamefont {A.}~\bibnamefont {Doostmohammadi}}, \bibinfo {author} {\bibfnamefont {T.~B.}\ \bibnamefont {Saw}}, \bibinfo {author} {\bibfnamefont {G.~H. N.~S.}\ \bibnamefont {Narayana}}, \bibinfo {author} {\bibfnamefont {R.}~\bibnamefont {Mueller}}, \bibinfo {author} {\bibfnamefont {T.}~\bibnamefont {Dang}}, \bibinfo {author} {\bibfnamefont {M.}~\bibnamefont {Thomas}}, \bibinfo {author} {\bibfnamefont {S.}~\bibnamefont {Gupta}}, \bibinfo {author} {\bibfnamefont {S.}~\bibnamefont {Sonam}}, \bibinfo {author} {\bibfnamefont {A.~S.}\ \bibnamefont {Yap}}, \bibinfo {author} {\bibfnamefont {Y.}~\bibnamefont {Toyama}}, \bibinfo {author} {\bibfnamefont {R.-M.}\ \bibnamefont {Mège}}, \bibinfo {author} {\bibfnamefont {J.~M.}\ \bibnamefont {Yeomans}},\ and\ \bibinfo {author} {\bibfnamefont {B.}~\bibnamefont {Ladoux}},\ }\bibfield  {title} {{\selectlanguage {en}\bibinfo {title} {Investigating the nature of
  active forces in tissues reveals how contractile cells can form extensile monolayers}},\ }\href {https://doi.org/10.1038/s41563-021-00919-2} {\bibfield  {journal} {\bibinfo  {journal} {Nature Materials}\ }\textbf {\bibinfo {volume} {20}},\ \bibinfo {pages} {1156} (\bibinfo {year} {2021})}\BibitemShut {NoStop}%
\bibitem [{\citenamefont {Giomi}\ \emph {et~al.}(2013)\citenamefont {Giomi}, \citenamefont {Bowick}, \citenamefont {Ma},\ and\ \citenamefont {Marchetti}}]{giomi_defect_2013}%
  \BibitemOpen
  \bibfield  {author} {\bibinfo {author} {\bibfnamefont {L.}~\bibnamefont {Giomi}}, \bibinfo {author} {\bibfnamefont {M.~J.}\ \bibnamefont {Bowick}}, \bibinfo {author} {\bibfnamefont {X.}~\bibnamefont {Ma}},\ and\ \bibinfo {author} {\bibfnamefont {M.~C.}\ \bibnamefont {Marchetti}},\ }\bibfield  {title} {\bibinfo {title} {Defect {Annihilation} and {Proliferation} in {Active} {Nematics}},\ }\href {https://doi.org/10.1103/PhysRevLett.110.228101} {\bibfield  {journal} {\bibinfo  {journal} {Physical Review Letters}\ }\textbf {\bibinfo {volume} {110}},\ \bibinfo {pages} {228101} (\bibinfo {year} {2013})}\BibitemShut {NoStop}%
\bibitem [{\citenamefont {Wioland}\ \emph {et~al.}(2016)\citenamefont {Wioland}, \citenamefont {Woodhouse}, \citenamefont {Dunkel},\ and\ \citenamefont {Goldstein}}]{wioland_ferromagnetic_2016}%
  \BibitemOpen
  \bibfield  {author} {\bibinfo {author} {\bibfnamefont {H.}~\bibnamefont {Wioland}}, \bibinfo {author} {\bibfnamefont {F.~G.}\ \bibnamefont {Woodhouse}}, \bibinfo {author} {\bibfnamefont {J.}~\bibnamefont {Dunkel}},\ and\ \bibinfo {author} {\bibfnamefont {R.~E.}\ \bibnamefont {Goldstein}},\ }\bibfield  {title} {{\selectlanguage {en}\bibinfo {title} {Ferromagnetic and antiferromagnetic order in bacterial vortex lattices}},\ }\href {https://doi.org/10.1038/nphys3607} {\bibfield  {journal} {\bibinfo  {journal} {Nature Physics}\ }\textbf {\bibinfo {volume} {12}},\ \bibinfo {pages} {341} (\bibinfo {year} {2016})}\BibitemShut {NoStop}%
\bibitem [{\citenamefont {Shankar}\ \emph {et~al.}(2022)\citenamefont {Shankar}, \citenamefont {Souslov}, \citenamefont {Bowick}, \citenamefont {Marchetti},\ and\ \citenamefont {Vitelli}}]{shankar_topological_2022}%
  \BibitemOpen
  \bibfield  {author} {\bibinfo {author} {\bibfnamefont {S.}~\bibnamefont {Shankar}}, \bibinfo {author} {\bibfnamefont {A.}~\bibnamefont {Souslov}}, \bibinfo {author} {\bibfnamefont {M.~J.}\ \bibnamefont {Bowick}}, \bibinfo {author} {\bibfnamefont {M.~C.}\ \bibnamefont {Marchetti}},\ and\ \bibinfo {author} {\bibfnamefont {V.}~\bibnamefont {Vitelli}},\ }\bibfield  {title} {{\selectlanguage {en}\bibinfo {title} {Topological active matter}},\ }\href {https://doi.org/10.1038/s42254-022-00445-3} {\bibfield  {journal} {\bibinfo  {journal} {Nature Reviews Physics}\ }\textbf {\bibinfo {volume} {4}},\ \bibinfo {pages} {380} (\bibinfo {year} {2022})}\BibitemShut {NoStop}%
\bibitem [{\citenamefont {{\v{C}}opar}\ \emph {et~al.}(2019)\citenamefont {{\v{C}}opar}, \citenamefont {Aplinc}, \citenamefont {Kos}, \citenamefont {{\v{Z}}umer},\ and\ \citenamefont {Ravnik}}]{vcopar2019topology}%
  \BibitemOpen
  \bibfield  {author} {\bibinfo {author} {\bibfnamefont {S.}~\bibnamefont {{\v{C}}opar}}, \bibinfo {author} {\bibfnamefont {J.}~\bibnamefont {Aplinc}}, \bibinfo {author} {\bibfnamefont {{\v{Z}}.}~\bibnamefont {Kos}}, \bibinfo {author} {\bibfnamefont {S.}~\bibnamefont {{\v{Z}}umer}},\ and\ \bibinfo {author} {\bibfnamefont {M.}~\bibnamefont {Ravnik}},\ }\bibfield  {title} {\bibinfo {title} {Topology of three-dimensional active nematic turbulence confined to droplets},\ }\href@noop {} {\bibfield  {journal} {\bibinfo  {journal} {Physical Review X}\ }\textbf {\bibinfo {volume} {9}},\ \bibinfo {pages} {031051} (\bibinfo {year} {2019})}\BibitemShut {NoStop}%
\bibitem [{\citenamefont {Keber}\ \emph {et~al.}(2014)\citenamefont {Keber}, \citenamefont {Loiseau}, \citenamefont {Sanchez}, \citenamefont {DeCamp}, \citenamefont {Giomi}, \citenamefont {Bowick}, \citenamefont {Marchetti}, \citenamefont {Dogic},\ and\ \citenamefont {Bausch}}]{keber_topology_2014}%
  \BibitemOpen
  \bibfield  {author} {\bibinfo {author} {\bibfnamefont {F.~C.}\ \bibnamefont {Keber}}, \bibinfo {author} {\bibfnamefont {E.}~\bibnamefont {Loiseau}}, \bibinfo {author} {\bibfnamefont {T.}~\bibnamefont {Sanchez}}, \bibinfo {author} {\bibfnamefont {S.~J.}\ \bibnamefont {DeCamp}}, \bibinfo {author} {\bibfnamefont {L.}~\bibnamefont {Giomi}}, \bibinfo {author} {\bibfnamefont {M.~J.}\ \bibnamefont {Bowick}}, \bibinfo {author} {\bibfnamefont {M.~C.}\ \bibnamefont {Marchetti}}, \bibinfo {author} {\bibfnamefont {Z.}~\bibnamefont {Dogic}},\ and\ \bibinfo {author} {\bibfnamefont {A.~R.}\ \bibnamefont {Bausch}},\ }\bibfield  {title} {\bibinfo {title} {Topology and dynamics of active nematic vesicles},\ }\href {https://doi.org/10.1126/science.1254784} {\bibfield  {journal} {\bibinfo  {journal} {Science}\ }\textbf {\bibinfo {volume} {345}},\ \bibinfo {pages} {1135} (\bibinfo {year} {2014})}\BibitemShut {NoStop}%
\bibitem [{\citenamefont {Shendruk}\ \emph {et~al.}(2017)\citenamefont {Shendruk}, \citenamefont {Doostmohammadi}, \citenamefont {Thijssen},\ and\ \citenamefont {Yeomans}}]{shendruk_dancing_2017}%
  \BibitemOpen
  \bibfield  {author} {\bibinfo {author} {\bibfnamefont {T.~N.}\ \bibnamefont {Shendruk}}, \bibinfo {author} {\bibfnamefont {A.}~\bibnamefont {Doostmohammadi}}, \bibinfo {author} {\bibfnamefont {K.}~\bibnamefont {Thijssen}},\ and\ \bibinfo {author} {\bibfnamefont {J.~M.}\ \bibnamefont {Yeomans}},\ }\bibfield  {title} {{\selectlanguage {en}\bibinfo {title} {Dancing disclinations in confined active nematics}},\ }\href {https://doi.org/10.1039/C6SM02310J} {\bibfield  {journal} {\bibinfo  {journal} {Soft Matter}\ }\textbf {\bibinfo {volume} {13}},\ \bibinfo {pages} {3853} (\bibinfo {year} {2017})},\ \bibinfo {note} {{\textasciicircum}}\BibitemShut {NoStop}%
\bibitem [{\citenamefont {Vitelli}\ and\ \citenamefont {Nelson}(2004)}]{vitelli_defect_2004}%
  \BibitemOpen
  \bibfield  {author} {\bibinfo {author} {\bibfnamefont {V.}~\bibnamefont {Vitelli}}\ and\ \bibinfo {author} {\bibfnamefont {D.~R.}\ \bibnamefont {Nelson}},\ }\bibfield  {title} {\bibinfo {title} {Defect generation and deconfinement on corrugated topographies},\ }\href {https://doi.org/10.1103/PhysRevE.70.051105} {\bibfield  {journal} {\bibinfo  {journal} {Physical Review E}\ }\textbf {\bibinfo {volume} {70}},\ \bibinfo {pages} {051105} (\bibinfo {year} {2004})}\BibitemShut {NoStop}%
\bibitem [{\citenamefont {Pearce}(2020)}]{pearce_defect_2020}%
  \BibitemOpen
  \bibfield  {author} {\bibinfo {author} {\bibfnamefont {D.~J.~G.}\ \bibnamefont {Pearce}},\ }\bibfield  {title} {{\selectlanguage {en}\bibinfo {title} {Defect order in active nematics on a curved surface}},\ }\href {https://doi.org/10.1088/1367-2630/ab91fd} {\bibfield  {journal} {\bibinfo  {journal} {New Journal of Physics}\ }\textbf {\bibinfo {volume} {22}},\ \bibinfo {pages} {063051} (\bibinfo {year} {2020})}\BibitemShut {NoStop}%
\bibitem [{\citenamefont {Hardoüin}\ \emph {et~al.}(2022)\citenamefont {Hardoüin}, \citenamefont {Doré}, \citenamefont {Laurent}, \citenamefont {Lopez-Leon}, \citenamefont {Ignés-Mullol},\ and\ \citenamefont {Sagués}}]{hardouin_active_2022}%
  \BibitemOpen
  \bibfield  {author} {\bibinfo {author} {\bibfnamefont {J.}~\bibnamefont {Hardoüin}}, \bibinfo {author} {\bibfnamefont {C.}~\bibnamefont {Doré}}, \bibinfo {author} {\bibfnamefont {J.}~\bibnamefont {Laurent}}, \bibinfo {author} {\bibfnamefont {T.}~\bibnamefont {Lopez-Leon}}, \bibinfo {author} {\bibfnamefont {J.}~\bibnamefont {Ignés-Mullol}},\ and\ \bibinfo {author} {\bibfnamefont {F.}~\bibnamefont {Sagués}},\ }\bibfield  {title} {{\selectlanguage {en}\bibinfo {title} {Active boundary layers in confined active nematics}},\ }\href {https://doi.org/10.1038/s41467-022-34336-z} {\bibfield  {journal} {\bibinfo  {journal} {Nature Communications}\ }\textbf {\bibinfo {volume} {13}},\ \bibinfo {pages} {6675} (\bibinfo {year} {2022})}\BibitemShut {NoStop}%
\bibitem [{\citenamefont {Doostmohammadi}\ \emph {et~al.}(2018)\citenamefont {Doostmohammadi}, \citenamefont {Ignés-Mullol}, \citenamefont {Yeomans},\ and\ \citenamefont {Sagués}}]{doostmohammadi_active_2018}%
  \BibitemOpen
  \bibfield  {author} {\bibinfo {author} {\bibfnamefont {A.}~\bibnamefont {Doostmohammadi}}, \bibinfo {author} {\bibfnamefont {J.}~\bibnamefont {Ignés-Mullol}}, \bibinfo {author} {\bibfnamefont {J.~M.}\ \bibnamefont {Yeomans}},\ and\ \bibinfo {author} {\bibfnamefont {F.}~\bibnamefont {Sagués}},\ }\bibfield  {title} {{\selectlanguage {en}\bibinfo {title} {Active nematics}},\ }\href {https://doi.org/10.1038/s41467-018-05666-8} {\bibfield  {journal} {\bibinfo  {journal} {Nature Communications}\ }\textbf {\bibinfo {volume} {9}},\ \bibinfo {pages} {3246} (\bibinfo {year} {2018})}\BibitemShut {NoStop}%
\bibitem [{\citenamefont {Walton}\ \emph {et~al.}(2020)\citenamefont {Walton}, \citenamefont {McKay}, \citenamefont {Grinfeld},\ and\ \citenamefont {Mottram}}]{walton_pressure-driven_2020}%
  \BibitemOpen
  \bibfield  {author} {\bibinfo {author} {\bibfnamefont {J.}~\bibnamefont {Walton}}, \bibinfo {author} {\bibfnamefont {G.}~\bibnamefont {McKay}}, \bibinfo {author} {\bibfnamefont {M.}~\bibnamefont {Grinfeld}},\ and\ \bibinfo {author} {\bibfnamefont {N.~J.}\ \bibnamefont {Mottram}},\ }\bibfield  {title} {{\selectlanguage {en}\bibinfo {title} {Pressure-driven changes to spontaneous flow in active nematic liquid crystals}},\ }\href {https://doi.org/10.1140/epje/i2020-11973-8} {\bibfield  {journal} {\bibinfo  {journal} {The European Physical Journal E}\ }\textbf {\bibinfo {volume} {43}},\ \bibinfo {pages} {51} (\bibinfo {year} {2020})}\BibitemShut {NoStop}%
\bibitem [{\citenamefont {Joshi}\ \emph {et~al.}(2022)\citenamefont {Joshi}, \citenamefont {Ray}, \citenamefont {Lemma}, \citenamefont {Varghese}, \citenamefont {Sharp}, \citenamefont {Dogic}, \citenamefont {Baskaran},\ and\ \citenamefont {Hagan}}]{joshi_data-driven_2022}%
  \BibitemOpen
  \bibfield  {author} {\bibinfo {author} {\bibfnamefont {C.}~\bibnamefont {Joshi}}, \bibinfo {author} {\bibfnamefont {S.}~\bibnamefont {Ray}}, \bibinfo {author} {\bibfnamefont {L.~M.}\ \bibnamefont {Lemma}}, \bibinfo {author} {\bibfnamefont {M.}~\bibnamefont {Varghese}}, \bibinfo {author} {\bibfnamefont {G.}~\bibnamefont {Sharp}}, \bibinfo {author} {\bibfnamefont {Z.}~\bibnamefont {Dogic}}, \bibinfo {author} {\bibfnamefont {A.}~\bibnamefont {Baskaran}},\ and\ \bibinfo {author} {\bibfnamefont {M.~F.}\ \bibnamefont {Hagan}},\ }\bibfield  {title} {\bibinfo {title} {Data-{Driven} {Discovery} of {Active} {Nematic} {Hydrodynamics}},\ }\href {https://doi.org/10.1103/PhysRevLett.129.258001} {\bibfield  {journal} {\bibinfo  {journal} {Physical Review Letters}\ }\textbf {\bibinfo {volume} {129}},\ \bibinfo {pages} {258001} (\bibinfo {year} {2022})}\BibitemShut {NoStop}%
\bibitem [{Note1()}]{Note1}%
  \BibitemOpen
  \bibinfo {note} {\textcolor {black}{Technically, another estimate is the source correction $r^\star (\protect \mathbf {x},t)$, but we have observed that it is rarely far from zero (strictly) inside the field of view.}}\BibitemShut {Stop}%
\bibitem [{\citenamefont {Boquet-Pujadas}\ and\ \citenamefont {Olivo-Marin}(2022)}]{boquet-pujadas_reformulating_2022}%
  \BibitemOpen
  \bibfield  {author} {\bibinfo {author} {\bibfnamefont {A.}~\bibnamefont {Boquet-Pujadas}}\ and\ \bibinfo {author} {\bibfnamefont {J.-C.}\ \bibnamefont {Olivo-Marin}},\ }\bibfield  {title} {\bibinfo {title} {Reformulating {Optical} {Flow} to {Solve} {Image}-{Based} {Inverse} {Problems} and {Quantify} {Uncertainty}},\ }\href {https://doi.org/10.1109/TPAMI.2022.3202855} {\bibfield  {journal} {\bibinfo  {journal} {IEEE Transactions on Pattern Analysis and Machine Intelligence}\ ,\ \bibinfo {pages} {1}} (\bibinfo {year} {2022})}\BibitemShut {NoStop}%
\bibitem [{Note2()}]{Note2}%
  \BibitemOpen
  \bibinfo {note} {In figures, we portray them using ``viridis'' and ``jet'' colormaps, respectively. The units and the ranges for the colorbars are always in the captions.}\BibitemShut {Stop}%
\bibitem [{\citenamefont {Hardoüin}\ \emph {et~al.}(2019)\citenamefont {Hardoüin}, \citenamefont {Hughes}, \citenamefont {Doostmohammadi}, \citenamefont {Laurent}, \citenamefont {Lopez-Leon}, \citenamefont {Yeomans}, \citenamefont {Ignés-Mullol},\ and\ \citenamefont {Sagués}}]{hardouin_reconfigurable_2019}%
  \BibitemOpen
  \bibfield  {author} {\bibinfo {author} {\bibfnamefont {J.}~\bibnamefont {Hardoüin}}, \bibinfo {author} {\bibfnamefont {R.}~\bibnamefont {Hughes}}, \bibinfo {author} {\bibfnamefont {A.}~\bibnamefont {Doostmohammadi}}, \bibinfo {author} {\bibfnamefont {J.}~\bibnamefont {Laurent}}, \bibinfo {author} {\bibfnamefont {T.}~\bibnamefont {Lopez-Leon}}, \bibinfo {author} {\bibfnamefont {J.~M.}\ \bibnamefont {Yeomans}}, \bibinfo {author} {\bibfnamefont {J.}~\bibnamefont {Ignés-Mullol}},\ and\ \bibinfo {author} {\bibfnamefont {F.}~\bibnamefont {Sagués}},\ }\bibfield  {title} {{\selectlanguage {en}\bibinfo {title} {Reconfigurable flows and defect landscape of confined active nematics}},\ }\href {https://doi.org/10.1038/s42005-019-0221-x} {\bibfield  {journal} {\bibinfo  {journal} {Communications Physics}\ }\textbf {\bibinfo {volume} {2}},\ \bibinfo {pages} {1} (\bibinfo {year} {2019})}\BibitemShut {NoStop}%
\bibitem [{\citenamefont {Shadden}\ \emph {et~al.}(2005)\citenamefont {Shadden}, \citenamefont {Lekien},\ and\ \citenamefont {Marsden}}]{shadden_definition_2005}%
  \BibitemOpen
  \bibfield  {author} {\bibinfo {author} {\bibfnamefont {S.~C.}\ \bibnamefont {Shadden}}, \bibinfo {author} {\bibfnamefont {F.}~\bibnamefont {Lekien}},\ and\ \bibinfo {author} {\bibfnamefont {J.~E.}\ \bibnamefont {Marsden}},\ }\bibfield  {title} {{\selectlanguage {en}\bibinfo {title} {Definition and properties of {Lagrangian} coherent structures from finite-time {Lyapunov} exponents in two-dimensional aperiodic flows}},\ }\href {https://doi.org/10.1016/j.physd.2005.10.007} {\bibfield  {journal} {\bibinfo  {journal} {Physica D: Nonlinear Phenomena}\ }\textbf {\bibinfo {volume} {212}},\ \bibinfo {pages} {271} (\bibinfo {year} {2005})}\BibitemShut {NoStop}%
\bibitem [{\citenamefont {Boquet~Pujadas}(2019)}]{boquet_pujadas_variational_2019}%
  \BibitemOpen
  \bibfield  {author} {\bibinfo {author} {\bibfnamefont {A.}~\bibnamefont {Boquet~Pujadas}},\ }\emph {\bibinfo {title} {Variational approaches in inverse problems for image-based characterisation of cellular dynamics}},\ \href {https://www.theses.fr/2019SORUS558} {\bibinfo {type} {Doctoral thesis}},\ \bibinfo  {school} {Sorbonne Université} (\bibinfo {year} {2019}),\ \bibinfo {note} {chapter II.5, pp. 98-107}\BibitemShut {NoStop}%
\bibitem [{\citenamefont {Serra}\ \emph {et~al.}(2023)\citenamefont {Serra}, \citenamefont {Lemma}, \citenamefont {Giomi}, \citenamefont {Dogic},\ and\ \citenamefont {Mahadevan}}]{serra_defect-mediated_2023}%
  \BibitemOpen
  \bibfield  {author} {\bibinfo {author} {\bibfnamefont {M.}~\bibnamefont {Serra}}, \bibinfo {author} {\bibfnamefont {L.}~\bibnamefont {Lemma}}, \bibinfo {author} {\bibfnamefont {L.}~\bibnamefont {Giomi}}, \bibinfo {author} {\bibfnamefont {Z.}~\bibnamefont {Dogic}},\ and\ \bibinfo {author} {\bibfnamefont {L.}~\bibnamefont {Mahadevan}},\ }\bibfield  {title} {{\selectlanguage {en}\bibinfo {title} {Defect-mediated dynamics of coherent structures in active nematics}},\ }\href {https://doi.org/10.1038/s41567-023-02062-y} {\bibfield  {journal} {\bibinfo  {journal} {Nature Physics}\ }\textbf {\bibinfo {volume} {19}},\ \bibinfo {pages} {1355} (\bibinfo {year} {2023})},\ \bibinfo {note} {publisher: Nature Publishing Group}\BibitemShut {NoStop}%
\bibitem [{\citenamefont {Giomi}\ \emph {et~al.}(2011)\citenamefont {Giomi}, \citenamefont {Mahadevan}, \citenamefont {Chakraborty},\ and\ \citenamefont {Hagan}}]{giomi_excitable_2011}%
  \BibitemOpen
  \bibfield  {author} {\bibinfo {author} {\bibfnamefont {L.}~\bibnamefont {Giomi}}, \bibinfo {author} {\bibfnamefont {L.}~\bibnamefont {Mahadevan}}, \bibinfo {author} {\bibfnamefont {B.}~\bibnamefont {Chakraborty}},\ and\ \bibinfo {author} {\bibfnamefont {M.~F.}\ \bibnamefont {Hagan}},\ }\bibfield  {title} {\bibinfo {title} {Excitable {Patterns} in {Active} {Nematics}},\ }\href {https://doi.org/10.1103/PhysRevLett.106.218101} {\bibfield  {journal} {\bibinfo  {journal} {Physical Review Letters}\ }\textbf {\bibinfo {volume} {106}},\ \bibinfo {pages} {218101} (\bibinfo {year} {2011})}\BibitemShut {NoStop}%
\bibitem [{\citenamefont {Shankar}\ and\ \citenamefont {Marchetti}(2019)}]{shankar2019hydrodynamics}%
  \BibitemOpen
  \bibfield  {author} {\bibinfo {author} {\bibfnamefont {S.}~\bibnamefont {Shankar}}\ and\ \bibinfo {author} {\bibfnamefont {M.~C.}\ \bibnamefont {Marchetti}},\ }\bibfield  {title} {\bibinfo {title} {Hydrodynamics of active defects: From order to chaos to defect ordering},\ }\href@noop {} {\bibfield  {journal} {\bibinfo  {journal} {Physical Review X}\ }\textbf {\bibinfo {volume} {9}},\ \bibinfo {pages} {041047} (\bibinfo {year} {2019})}\BibitemShut {NoStop}%
\bibitem [{\citenamefont {Thijssen}\ \emph {et~al.}(2021)\citenamefont {Thijssen}, \citenamefont {Khaladj}, \citenamefont {Aghvami}, \citenamefont {Gharbi}, \citenamefont {Fraden}, \citenamefont {Yeomans}, \citenamefont {Hirst},\ and\ \citenamefont {Shendruk}}]{thijssen_submersed_2021}%
  \BibitemOpen
  \bibfield  {author} {\bibinfo {author} {\bibfnamefont {K.}~\bibnamefont {Thijssen}}, \bibinfo {author} {\bibfnamefont {D.~A.}\ \bibnamefont {Khaladj}}, \bibinfo {author} {\bibfnamefont {S.~A.}\ \bibnamefont {Aghvami}}, \bibinfo {author} {\bibfnamefont {M.~A.}\ \bibnamefont {Gharbi}}, \bibinfo {author} {\bibfnamefont {S.}~\bibnamefont {Fraden}}, \bibinfo {author} {\bibfnamefont {J.~M.}\ \bibnamefont {Yeomans}}, \bibinfo {author} {\bibfnamefont {L.~S.}\ \bibnamefont {Hirst}},\ and\ \bibinfo {author} {\bibfnamefont {T.~N.}\ \bibnamefont {Shendruk}},\ }\bibfield  {title} {{\selectlanguage {en}\bibinfo {title} {Submersed micropatterned structures control active nematic flow, topology, and concentration}},\ }\href {https://doi.org/10.1073/pnas.2106038118} {\bibfield  {journal} {\bibinfo  {journal} {Proceedings of the National Academy of Sciences}\ }\textbf {\bibinfo {volume} {118}},\ \bibinfo {pages} {e2106038118} (\bibinfo {year} {2021})}\BibitemShut {NoStop}%
\bibitem [{Note3()}]{Note3}%
  \BibitemOpen
  \bibinfo {note} {As opposed to forward problems (simulations), inverse problems try to guess the cause from the effect using incomplete or noisy data. For example, magnetic-resonance imaging reconstructs the concentration of hydrogen from radio-frequency signals, and geologists probe the Earth's mantle with seismographs. Other inverse problems are computed tomography and ultrasound imaging. Our inverse problem is conceptually similar. The main differences are twofold. First, standard inverse problems fit the data directly with a least-squares error. Instead, here we use a conservation equation for the data term. And second, we use a PDE-constraint instead of the standard system matrix.}\BibitemShut {Stop}%
\bibitem [{\citenamefont {Morozov}(1984)}]{morozov_criteria_1984}%
  \BibitemOpen
  \bibfield  {author} {\bibinfo {author} {\bibfnamefont {V.~A.}\ \bibnamefont {Morozov}},\ }\bibfield  {title} {\bibinfo {title} {Criteria for {Selection} of {Regularization} {Parameter}},\ }in\ \href {https://doi.org/10.1007/978-1-4612-5280-1_2} {\emph {\bibinfo {booktitle} {Methods for {Solving} {Incorrectly} {Posed} {Problems}}}},\ \bibinfo {editor} {edited by\ \bibinfo {editor} {\bibfnamefont {V.~A.}\ \bibnamefont {Morozov}}}\ (\bibinfo  {publisher} {Springer},\ \bibinfo {address} {New York, NY},\ \bibinfo {year} {1984})\ pp.\ \bibinfo {pages} {32--64}\BibitemShut {NoStop}%
\bibitem [{\citenamefont {Salençon}(2001)}]{salencon_virtual_2001}%
  \BibitemOpen
  \bibfield  {author} {\bibinfo {author} {\bibfnamefont {J.}~\bibnamefont {Salençon}},\ }\bibfield  {title} {{\selectlanguage {en}\bibinfo {title} {The {Virtual} {Work} {Approach} to the {Modelling} of {Forces}}},\ }in\ \href {https://doi.org/10.1007/978-3-642-56542-7_4} {{\selectlanguage {en}\emph {\bibinfo {booktitle} {Handbook of {Continuum} {Mechanics}: {General} {Concepts} {Thermoelasticity}}}}},\ \bibinfo {editor} {edited by\ \bibinfo {editor} {\bibfnamefont {J.}~\bibnamefont {Salençon}}}\ (\bibinfo  {publisher} {Springer},\ \bibinfo {address} {Berlin, Heidelberg},\ \bibinfo {year} {2001})\ pp.\ \bibinfo {pages} {133--181}\BibitemShut {NoStop}%
\bibitem [{\citenamefont {Boquet-Pujadas}\ \emph {et~al.}(2022)\citenamefont {Boquet-Pujadas}, \citenamefont {Feaugas}, \citenamefont {Petracchini}, \citenamefont {Grassart}, \citenamefont {Mary}, \citenamefont {Manich}, \citenamefont {Gobaa}, \citenamefont {Olivo-Marin}, \citenamefont {Sauvonnet},\ and\ \citenamefont {Labruyère}}]{boquet-pujadas_4d_2022}%
  \BibitemOpen
  \bibfield  {author} {\bibinfo {author} {\bibfnamefont {A.}~\bibnamefont {Boquet-Pujadas}}, \bibinfo {author} {\bibfnamefont {T.}~\bibnamefont {Feaugas}}, \bibinfo {author} {\bibfnamefont {A.}~\bibnamefont {Petracchini}}, \bibinfo {author} {\bibfnamefont {A.}~\bibnamefont {Grassart}}, \bibinfo {author} {\bibfnamefont {H.}~\bibnamefont {Mary}}, \bibinfo {author} {\bibfnamefont {M.}~\bibnamefont {Manich}}, \bibinfo {author} {\bibfnamefont {S.}~\bibnamefont {Gobaa}}, \bibinfo {author} {\bibfnamefont {J.-C.}\ \bibnamefont {Olivo-Marin}}, \bibinfo {author} {\bibfnamefont {N.}~\bibnamefont {Sauvonnet}},\ and\ \bibinfo {author} {\bibfnamefont {E.}~\bibnamefont {Labruyère}},\ }\bibfield  {title} {\bibinfo {title} {{4D} live imaging and computational modeling of a functional gut-on-a-chip evaluate how peristalsis facilitates enteric pathogen invasion},\ }\href {https://doi.org/10.1126/sciadv.abo5767} {\bibfield  {journal} {\bibinfo  {journal} {Science Advances}\ }\textbf {\bibinfo {volume} {8}},\ \bibinfo {pages}
  {eabo5767} (\bibinfo {year} {2022})}\BibitemShut {NoStop}%
\bibitem [{\citenamefont {Chambolle}\ \emph {et~al.}(2010)\citenamefont {Chambolle}, \citenamefont {Caselles}, \citenamefont {Cremers}, \citenamefont {Novaga},\ and\ \citenamefont {Pock}}]{chambolle_introduction_2010}%
  \BibitemOpen
  \bibfield  {author} {\bibinfo {author} {\bibfnamefont {A.}~\bibnamefont {Chambolle}}, \bibinfo {author} {\bibfnamefont {V.}~\bibnamefont {Caselles}}, \bibinfo {author} {\bibfnamefont {D.}~\bibnamefont {Cremers}}, \bibinfo {author} {\bibfnamefont {M.}~\bibnamefont {Novaga}},\ and\ \bibinfo {author} {\bibfnamefont {T.}~\bibnamefont {Pock}},\ }\bibfield  {title} {{\selectlanguage {en}\bibinfo {title} {An {Introduction} to {Total} {Variation} for {Image} {Analysis}}},\ }in\ \href {https://www.degruyter.com/document/doi/10.1515/9783110226157.263/html} {{\selectlanguage {en}\emph {\bibinfo {booktitle} {An {Introduction} to {Total} {Variation} for {Image} {Analysis}}}}}\ (\bibinfo  {publisher} {De Gruyter},\ \bibinfo {year} {2010})\ pp.\ \bibinfo {pages} {263--340}\BibitemShut {NoStop}%
\bibitem [{\citenamefont {Béréziat}\ and\ \citenamefont {Herlin}(2011)}]{bereziat_solving_2011}%
  \BibitemOpen
  \bibfield  {author} {\bibinfo {author} {\bibfnamefont {D.}~\bibnamefont {Béréziat}}\ and\ \bibinfo {author} {\bibfnamefont {I.}~\bibnamefont {Herlin}},\ }\bibfield  {title} {{\selectlanguage {en}\bibinfo {title} {Solving ill-posed {Image} {Processing} problems using {Data} {Assimilation}}},\ }\href {https://doi.org/10.1007/s11075-010-9383-z} {\bibfield  {journal} {\bibinfo  {journal} {Numerical Algorithms}\ }\textbf {\bibinfo {volume} {56}},\ \bibinfo {pages} {219} (\bibinfo {year} {2011})}\BibitemShut {NoStop}%
\bibitem [{\citenamefont {Héas}\ \emph {et~al.}(2016)\citenamefont {Héas}, \citenamefont {Drémeau},\ and\ \citenamefont {Herzet}}]{heas_efficient_2016}%
  \BibitemOpen
  \bibfield  {author} {\bibinfo {author} {\bibfnamefont {P.}~\bibnamefont {Héas}}, \bibinfo {author} {\bibfnamefont {A.}~\bibnamefont {Drémeau}},\ and\ \bibinfo {author} {\bibfnamefont {C.}~\bibnamefont {Herzet}},\ }\bibfield  {title} {\bibinfo {title} {An {Efficient} {Algorithm} for {Video} {Superresolution} {Based} on a {Sequential} {Model}},\ }\href {https://doi.org/10.1137/15M1023956} {\bibfield  {journal} {\bibinfo  {journal} {SIAM Journal on Imaging Sciences}\ }\textbf {\bibinfo {volume} {9}},\ \bibinfo {pages} {537} (\bibinfo {year} {2016})}\BibitemShut {NoStop}%
\bibitem [{\citenamefont {Alnæs}\ \emph {et~al.}(2015)\citenamefont {Alnæs}, \citenamefont {Blechta}, \citenamefont {Hake}, \citenamefont {Johansson}, \citenamefont {Kehlet}, \citenamefont {Logg}, \citenamefont {Richardson}, \citenamefont {Ring}, \citenamefont {Rognes},\ and\ \citenamefont {Wells}}]{alnaes_fenics_2015}%
  \BibitemOpen
  \bibfield  {author} {\bibinfo {author} {\bibfnamefont {M.}~\bibnamefont {Alnæs}}, \bibinfo {author} {\bibfnamefont {J.}~\bibnamefont {Blechta}}, \bibinfo {author} {\bibfnamefont {J.}~\bibnamefont {Hake}}, \bibinfo {author} {\bibfnamefont {A.}~\bibnamefont {Johansson}}, \bibinfo {author} {\bibfnamefont {B.}~\bibnamefont {Kehlet}}, \bibinfo {author} {\bibfnamefont {A.}~\bibnamefont {Logg}}, \bibinfo {author} {\bibfnamefont {C.}~\bibnamefont {Richardson}}, \bibinfo {author} {\bibfnamefont {J.}~\bibnamefont {Ring}}, \bibinfo {author} {\bibfnamefont {M.~E.}\ \bibnamefont {Rognes}},\ and\ \bibinfo {author} {\bibfnamefont {G.~N.}\ \bibnamefont {Wells}},\ }\bibfield  {title} {{\selectlanguage {en}\bibinfo {title} {The {FEniCS} {Project} {Version} 1.5}},\ }\bibfield  {journal} {\bibinfo  {journal} {Archive of Numerical Software}\ }\textbf {\bibinfo {volume} {3}},\ \href {https://doi.org/10.11588/ans.2015.100.20553} {10.11588/ans.2015.100.20553} (\bibinfo {year} {2015}),\ \bibinfo {note} {number: 100}\BibitemShut
  {NoStop}%
\bibitem [{\citenamefont {Vélez-Cerón}\ \emph {et~al.}(2024)\citenamefont {Vélez-Cerón}, \citenamefont {Guillamat}, \citenamefont {Sagués},\ and\ \citenamefont {Ignés-Mullol}}]{velez-ceron_probing_2024}%
  \BibitemOpen
  \bibfield  {author} {\bibinfo {author} {\bibfnamefont {I.}~\bibnamefont {Vélez-Cerón}}, \bibinfo {author} {\bibfnamefont {P.}~\bibnamefont {Guillamat}}, \bibinfo {author} {\bibfnamefont {F.}~\bibnamefont {Sagués}},\ and\ \bibinfo {author} {\bibfnamefont {J.}~\bibnamefont {Ignés-Mullol}},\ }\bibfield  {title} {\bibinfo {title} {Probing active nematics with in situ microfabricated elastic inclusions},\ }\href {https://doi.org/10.1073/pnas.2312494121} {\bibfield  {journal} {\bibinfo  {journal} {Proceedings of the National Academy of Sciences}\ }\textbf {\bibinfo {volume} {121}},\ \bibinfo {pages} {e2312494121} (\bibinfo {year} {2024})}\BibitemShut {NoStop}%
\bibitem [{\citenamefont {Selim}(2012)}]{selim_adaptive_2012}%
  \BibitemOpen
  \bibfield  {author} {\bibinfo {author} {\bibfnamefont {K.}~\bibnamefont {Selim}},\ }\bibfield  {title} {{\selectlanguage {en}\bibinfo {title} {An adaptive finite element solver for fluid–structure interaction problems}},\ }in\ \href {https://doi.org/10.1007/978-3-642-23099-8_29} {{\selectlanguage {en}\emph {\bibinfo {booktitle} {Automated {Solution} of {Differential} {Equations} by the {Finite} {Element} {Method}: {The} {FEniCS} {Book}}}}},\ \bibinfo {editor} {edited by\ \bibinfo {editor} {\bibfnamefont {A.}~\bibnamefont {Logg}}, \bibinfo {editor} {\bibfnamefont {K.-A.}\ \bibnamefont {Mardal}},\ and\ \bibinfo {editor} {\bibfnamefont {G.}~\bibnamefont {Wells}}}\ (\bibinfo  {publisher} {Springer},\ \bibinfo {address} {Berlin, Heidelberg},\ \bibinfo {year} {2012})\ pp.\ \bibinfo {pages} {553--569}\BibitemShut {NoStop}%
\bibitem [{Note4()}]{Note4}%
  \BibitemOpen
  \bibinfo {note} {Otherwise, one can perform the integral or use more specific pre-computed cantilever formulae.}\BibitemShut {Stop}%
\bibitem [{\citenamefont {Lee}\ \emph {et~al.}(2008)\citenamefont {Lee}, \citenamefont {Wei}, \citenamefont {Kysar},\ and\ \citenamefont {Hone}}]{lee_measurement_2008}%
  \BibitemOpen
  \bibfield  {author} {\bibinfo {author} {\bibfnamefont {C.}~\bibnamefont {Lee}}, \bibinfo {author} {\bibfnamefont {X.}~\bibnamefont {Wei}}, \bibinfo {author} {\bibfnamefont {J.~W.}\ \bibnamefont {Kysar}},\ and\ \bibinfo {author} {\bibfnamefont {J.}~\bibnamefont {Hone}},\ }\bibfield  {title} {\bibinfo {title} {Measurement of the {Elastic} {Properties} and {Intrinsic} {Strength} of {Monolayer} {Graphene}},\ }\href {https://doi.org/10.1126/science.1157996} {\bibfield  {journal} {\bibinfo  {journal} {Science}\ }\textbf {\bibinfo {volume} {321}},\ \bibinfo {pages} {385} (\bibinfo {year} {2008})},\ \bibinfo {note} {publisher: American Association for the Advancement of Science}\BibitemShut {NoStop}%
\bibitem [{\citenamefont {Thorpe}\ and\ \citenamefont {Jasiuk}(1992)}]{thorpe_new_1992}%
  \BibitemOpen
  \bibfield  {author} {\bibinfo {author} {\bibfnamefont {M.~F.}\ \bibnamefont {Thorpe}}\ and\ \bibinfo {author} {\bibfnamefont {I.}~\bibnamefont {Jasiuk}},\ }\bibfield  {title} {\bibinfo {title} {New {Results} in the {Theory} of {Elasticity} for {Two}-{Dimensional} {Composites}},\ }\href {https://www.jstor.org/stable/52148} {\bibfield  {journal} {\bibinfo  {journal} {Proceedings: Mathematical and Physical Sciences}\ }\textbf {\bibinfo {volume} {438}},\ \bibinfo {pages} {531} (\bibinfo {year} {1992})}\BibitemShut {NoStop}%
\bibitem [{\citenamefont {Guillamat}\ \emph {et~al.}(2016)\citenamefont {Guillamat}, \citenamefont {Ignés-Mullol},\ and\ \citenamefont {Sagués}}]{guillamat_control_2016}%
  \BibitemOpen
  \bibfield  {author} {\bibinfo {author} {\bibfnamefont {P.}~\bibnamefont {Guillamat}}, \bibinfo {author} {\bibfnamefont {J.}~\bibnamefont {Ignés-Mullol}},\ and\ \bibinfo {author} {\bibfnamefont {F.}~\bibnamefont {Sagués}},\ }\bibfield  {title} {\bibinfo {title} {Control of active liquid crystals with a magnetic field},\ }\href {https://doi.org/10.1073/pnas.1600339113} {\bibfield  {journal} {\bibinfo  {journal} {Proceedings of the National Academy of Sciences}\ }\textbf {\bibinfo {volume} {113}},\ \bibinfo {pages} {5498} (\bibinfo {year} {2016})}\BibitemShut {NoStop}%
\bibitem [{sup()}]{supp}%
  \BibitemOpen
  \href@noop {} {}\bibinfo {note} {See Supplemental Material at URL-will-be-inserted-by-publisher for the Movies.}\BibitemShut {Stop}%
\end{thebibliography}%

\end{document}